\documentclass[a4paper,11pt]{article}

\usepackage{setup}
\usepackage{jheppub}
\usepackage{todonotes}
\usepackage{placeins}

\usepackage{cite}
\usepackage{ifthen}
\usepackage{subcaption}
\usepackage{siunitx}
\usepackage{enumitem}
\usepackage{booktabs}
\usepackage{bookmark}

\usepackage{cleveref}
\DeclareSIUnit\fb{\femto\barn}

\renewcommand{\tabcolsep}{1em}

\def\lapprox{\lower .7ex\hbox{$\;\stackrel{\textstyle <}{\sim}\;$}}
\def\gapprox{\lower .7ex\hbox{$\;\stackrel{\textstyle >}{\sim}\;$}}
\definecolor{lightgray}{HTML}{A6A39A}
\definecolor{darkgray}{HTML}{504E48}
\definecolor{silver}{HTML}{E0DFDE}
\definecolor{brown}{HTML}{5F4541}
\definecolor{beige}{HTML}{DCCCAC}
\definecolor{green}{HTML}{345F53}
\definecolor{yellow}{HTML}{F6B65A}
\definecolor{blue}{HTML}{568BCF}
\definecolor{red}{HTML}{AE1932}
\definecolor{orange}{HTML}{D16F15}

\makeatletter
\newcommand{\myitem}[1]{%
	\item[#1]\protected@edef\@currentlabel{#1}%
}
\makeatother

\preprint{{\raggedleft%
ZU-TH 05/22 \\
}}

\title{QCD Predictions for Event-Shape Distributions in Hadronic Higgs Decays}
\author[a]{G.~Coloretti,}
\author[a,b]{A.~Gehrmann--De Ridder,}
\author[a]{C.~T.~Preuss}

\affiliation[a]{Institute for Theoretical Physics, ETH, CH-8093 Z\"urich, Switzerland}
\affiliation[b]{Department of Physics, University of Z\"urich, CH-8057 Z\"urich, Switzerland}

\emailAdd{guglielmo.coloretti@math.ethz.ch}
\emailAdd{gehra@phys.ethz.ch}
\emailAdd{cpreuss@phys.ethz.ch}

\abstract{
  We study the six classical event-shape observables in hadronic Higgs decays at next-to-leading order in QCD.
  To this end, we consider the decay of on-shell Higgs bosons to three partons, taking into account both
  the Yukawa-induced decay to $\Pqb$-quark pairs and the loop-induced decay to two gluons via
  an effective Higgs-gluon coupling.
  The results are discussed with a particular focus on the discriminative power of event shapes regarding these two classes 
  of processes.
}

\begin{document}
\maketitle
\flushbottom

\section{Introduction}
\label{sec:intro}
The Higgs boson was discovered by the ATLAS and CMS experiments at CERN in 2012 \cite{ATLAS:2012yve,CMS:2012qbp}. Since then, a major goal of present and future collider-physics programmes is the precise determination of Higgs-boson properties and the in-depth validation of the Higgs mechanism of electroweak symmetry breaking.
For the latter, one needs to firmly establish that the masses of all fundamental particles are generated through their interaction with the Higgs boson through precision measurements of each particle's coupling strength.
Up to now, the coupling of the Higgs boson to the electroweak gauge bosons $\PZ$ and $\PW^\pm$, as well as to third-generation 
fermions (bottom and top quarks, tau leptons) have been established at the LHC, and the associated coupling strengths were measured.
Within uncertainties, the properties and couplings of the Higgs boson are consistent with the predictions of the Standard Model of particle physics.
Detailed and comprehensive reports on the status of analyses of current and expected LHC data pertaining to the Higgs boson can be found in \cite{deFlorian:2016spz,Cepeda:2019klc}.

While the upcoming high-luminosity phase of the LHC (HL-LHC) will improve upon many of these measurements, some properties will remain elusive. Several Higgs-coupling measurements will be limited to accuracies around ten percent or above by the complexity related to the associated final states in hadronic collisions and their background processes. This accuracy is neither sufficient for real precision tests of the Higgs mechanism nor for testing scenarios for physics beyond the Standard Model, where one often expects deviations in the Higgs couplings at the level of at most few percents. 

In-depth precision studies of the Higgs boson and the associated Higgs mechanism will become possible at a future electron-positron collider that operates at a collision energy at or above $240~\GeV$, see  ~\cite{Abada:2019zxq,Moortgat-Picka:2015yla} for details. The currently investigated CERN FCC-ee~\cite{Abada:2019zxq}, CEPC~\cite{CEPCStudyGroup:2018ghi}, and ILC~\cite{Baer:2013cma} projects are all aiming to operate as Higgs factories, \ie, to produce a very large number of Higgs bosons for precision studies of its properties and couplings.

The striking advantage of $\Pep\Pem$ colliders comes from their clean experimental environment, where interactions take place at well-defined centre-of-mass energies and background processes are significantly suppressed compared to hadron-collider environments.

A future $\Pep\Pem$ collider will in particular enable model-independent measurements of the Higgs coupling to gauge bosons and fermions at the level of a few percent. It is expected that it will become possible to measure all decay channels of the Higgs boson, allowing for the precise determination of the total Higgs width.
This is in stark contrast to the current situation at hadron colliders where only specific decay channels 
can be identified and reconstructed reliably. 

The $\PH\to\Pqb\Paqb$ decay was observed at the LHC by the ATLAS~\cite{Aaboud:2018zhk} and CMS~\cite{Sirunyan:2018kst} collaborations in the $\PV\PH$ (${\rm V}=\PZ,\PW^\pm$) production modes with a significance of \numlist{5.3;5.6} standard deviations, respectively.
This decay is extremely important to measure, as it not only provides a direct measurement of the Higgs coupling to fermions, but also constitutes the dominant contribution to the total Higgs width. While it is difficult to measure hadronic Higgs decays in inclusive Higgs production via the leading gluon-fusion and vector-boson-fusion production modes at the LHC, the measurement of the $\PH\to\Pqb\Paqb$ decay was facilitated by the presence of a leptonically decaying vector boson in the $\PV\PH$ processes, which provides a clean experimental signature.
A similar approach to measure the sub-dominant $\PH\to\Pg\Pg$ decay at the LHC is, on the other hand, hindered by the overwhelming QCD background, despite impressive progress in distinguishing jets stemming from the Higgs decay from those originating from QCD background, see for example~\cite{Gras:2017jty,Mo:2017gzp,Cavallini:2021vot,Dreyer:2021hhr,Fedkevych:2022mid}.

Hadronic Higgs decays mainly proceed via two decay modes: either as a Yukawa-induced decay to a bottom-quark pair or as a loop-induced decay to two gluons.
Inclusive hadronic decay widths of the Higgs are theoretically known up to fourth order in the strong coupling \cite{Gorishnii:1990zu,Gorishnii:1991zr,Kataev:1993be,Surguladze:1994gc,Larin:1995sq,Chetyrkin:1995pd,Chetyrkin:1996sr,Baikov:2005rw} in the (massless) $\Pqb$-quark channel and up to third order in the di-gluon channel \cite{Spira:1995rr,Inami:1982xt,Chetyrkin:1997iv,Baikov:2006ch} under the assumption of an infinitely heavy top quark, with first-order electroweak corrections known in both cases \cite{Fleischer:1980ub,Bardin:1990zj,Dabelstein:1991ky,Kniehl:1991ze,Aglietti:2004nj,Degrassi:2004mx,Actis:2008ug,Aglietti:2004ki}.
Fully-differential second-order calculations, with massless b-quarks, have been performed for both the production and the decay of the Higgs into a bottom-quark pair~\cite{Ferrera:2017zex,Caola:2017xuq,Gauld:2019yng}, as well as the combination with resummation~\cite{Bizon:2019tfo} or parton showers~\cite{Alioli:2020fzf}.
In addition, the fully-differential $\PH\to\Pqb\Paqb$ decay rate has recently been computed at third order in the strong coupling~\cite{Mondini:2019gid}, however without flavour identification, on which the other fixed-order calculations heavily relied.

We here wish to focus on global event-shape distributions related to hadronic Higgs decays, which can be regarded as a class of ``good observables''.
On the experimental side, event shapes can be reconstructed from hadron momenta without the need to use a jet algorithm. 
Those can directly access the geometrical event properties, with different event-shapes variables probing different aspects of the underlying parton-level dynamics.
On the theory side, event shapes can be calculated reliably order-by-order in perturbative QCD and predictions can be compared faithfully with data away from the two-jet limit where resummation effects need to be taken into account.
At LEP, global event-shape distributions were studied extensively in $\PZ/\Pgg$ decays as a vital tool for QCD precision measurements, as for instance for the determination of the strong coupling ~\cite{Dissertori:2009qa, Dissertori:2009ik}
using the perturbative computations from \cite{Gehrmann-DeRidder:2007vsv,Gehrmann-DeRidder:2009fgd}.

To the best of our knowledge, comparable precision calculations of event-shape observables for hadronic Higgs decays have not been made public so far.
It is the purpose of this work to provide a comprehensive study of event-shapes in hadronic Higgs decays by computing the six classical  event shapes:  thrust, $C$-parameter, total and wide jet broadening, heavy-jet mass, and Durham three-jet resolution, at next-to-leading order (\NLO) in QCD.

As a practical application of our work, we wish to study the suitability of event-shape observables as discriminators between both hadronic Higgs-decay modes, as a means to facilitate the extraction of the involved Higgs couplings.
A similar idea has so-far only been pursued for the thrust observable, using either a three-jet merged calculation in \cite{Gao:2016jcm} or an \NLO calculation with approximate next-to-next-to-leading order (\NNLO) effects in \cite{Gao:2019mlt}, and for the energy-energy correlation, which has been computed at \NLO in \cite{Luo:2019nig, Gao:2020vyx}.

The structure of the paper is as follows: in \cref{sec:Framework}, we summarise the general framework of the calculation and provide details on all ingredients entering the computation of differential Higgs decays in perturbative QCD. Our results are presented in \cref{sec:results} and summarised in \cref{sec:conclusion}, where an outlook on future work is given as well.

\section{Differential Higgs decay rates at \NLO in QCD}
\label{sec:Framework}
In this work, we compute differential Higgs decay rates in the limit of vanishing light-quark masses, while keeping a non-vanishing Yukawa coupling only for the bottom quark and an infinitely large top-quark mass. As a consequence, the hadronic decay of the Higgs boson can be classified at parton level into two categories.

\begin{figure}[t]
  \centering
  \includegraphics[width=0.9\textwidth]{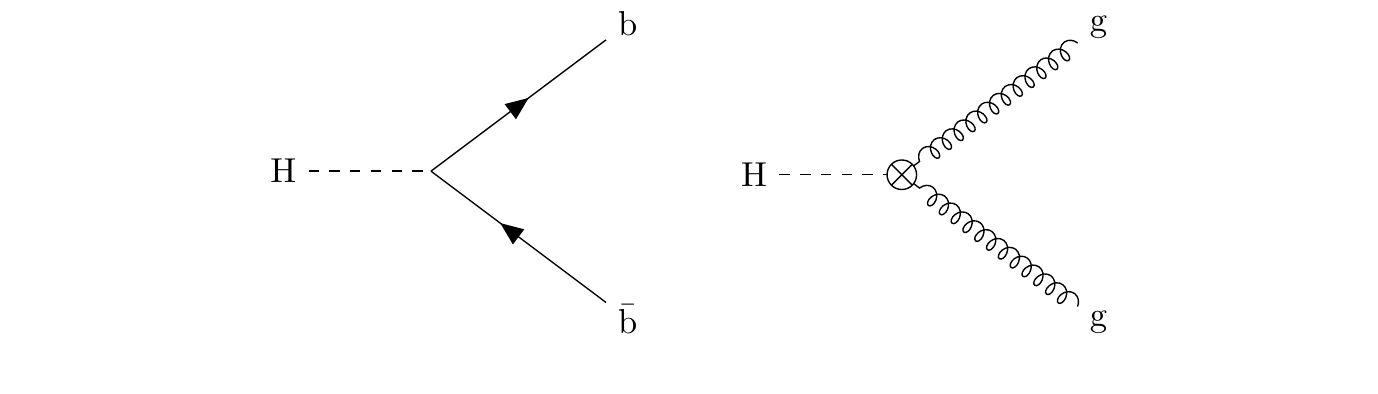}
  \caption{Hadronic Higgs decay categories: $\PH\to\Pqb\Paqb$ with a Yukawa coupling  (\textit{left}) and $\PH\to\Pg\Pg$ via an effective coupling (\textit{right}).}
  \label{fig:diagH2jLO}
\end{figure}

\begin{figure}[t]
  \centering
    \includegraphics[width=0.9\textwidth]{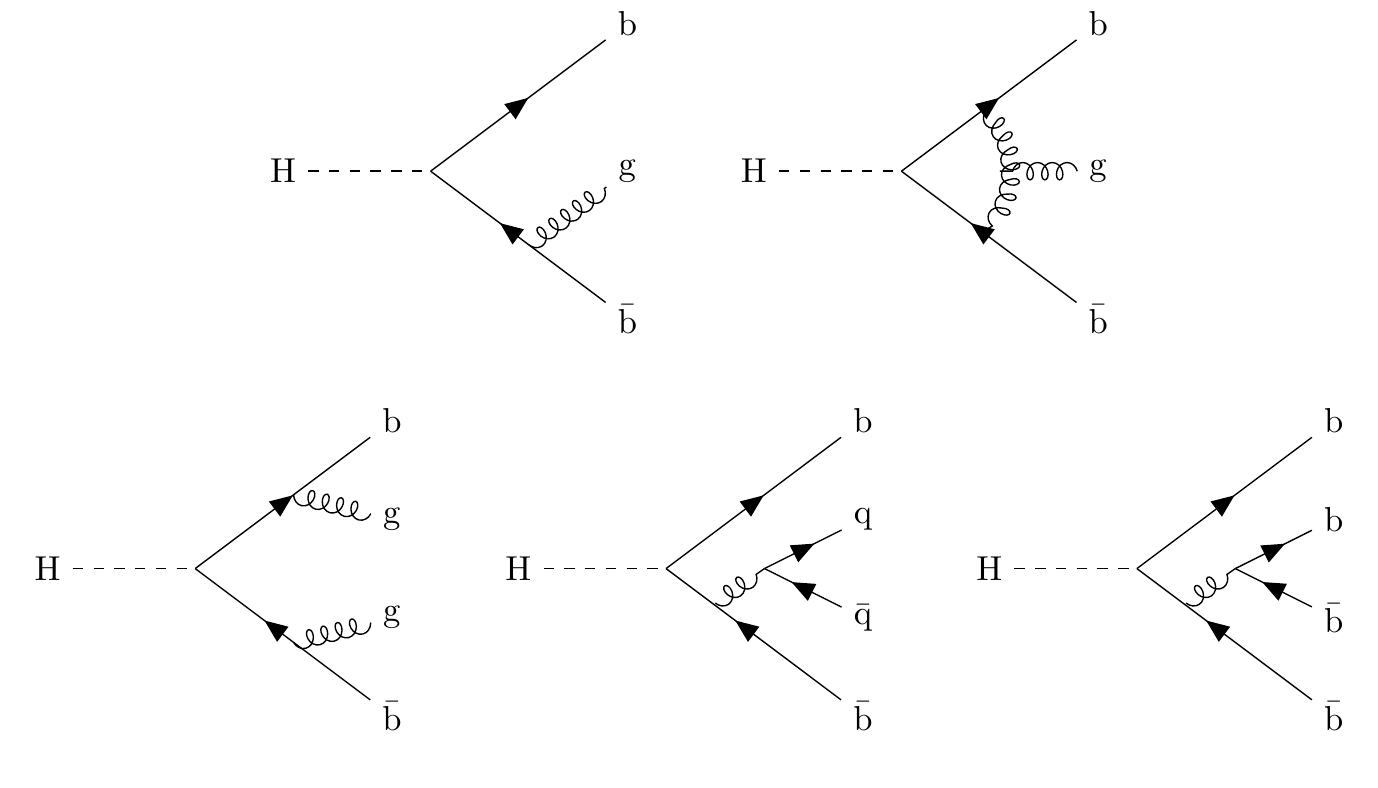}
  \caption{Representative Feynman diagrams for hadronic Higgs decays to bottom quarks  in the $\PH\to\Pqb\Paqb$ category up to second order in $\alphas$}
  \label{fig:diagH3j_bb}
\end{figure}

In the first category, the Higgs decay is associated to the Yukawa-type coupling $\yb$, and the decay can be obtained using the Standard-Model QCD Lagrangian.
Being generated by a primary $\PH\Pqb\Paqb$ vertex, contributions of this type will be referred to as belonging to the $\PH\to\Pqb\Paqb$ category in the remainder of this paper.
The associated two-parton decay diagram is shown in the left-hand panel of \cref{fig:diagH2jLO}.
Further, the corresponding three- and four-parton processes entering our calculation at \LO and \NLO are presented in the first column of \cref{tab:channels}, while representative Feynman diagrams at tree and one-loop level are displayed in \cref{fig:diagH3j_bb}.

\begin{figure}[ht]
  \centering
   \includegraphics[width=0.9\textwidth]{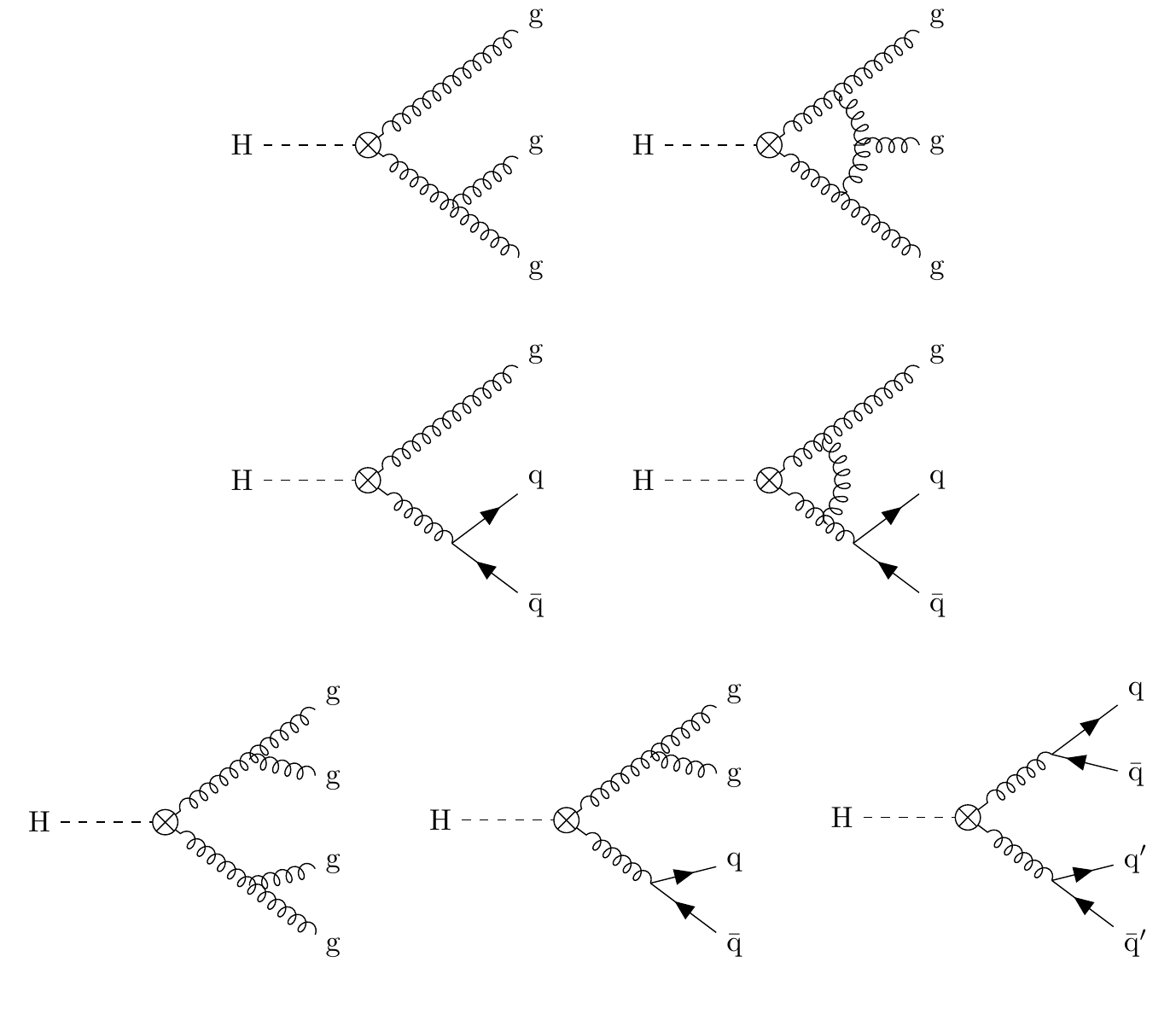}
 
  \caption{Representative Feynman diagrams for hadronic Higgs decays to gluons in the 
   $\PH\to\Pg\Pg$ category up to second order in $\alphas$. }
  \label{fig:diagH3j_gg}
\end{figure}

The coupling to quarks also enables the Higgs boson to couple to gluons via a heavy-quark loop.
In the limit of infinitely heavy top quarks, these top-quark loops decouple, yielding an effective theory containing a direct interaction of the Higgs field with the gluon field-strength tensor.
In this second category, the interaction is mediated by an effective $\PH\Pg\Pg$ vertex, represented as a crossed dot in the right-hand panel of \cref{fig:diagH2jLO}. 
We shall refer to this type of contributions as belonging to the $\PH\to\Pg\Pg$ category.
It is worth mentioning that in this category there are two distinct tree-level processes contributing to the three-parton rate.
A three-parton state can either be obtained via the emission of an additional gluon from the $\PH\Pg\Pg$ effective vertex or through the splitting of one of the primary gluons to a quark-antiquark pair.
A summary of the three- and four-parton processes contributing to this decay category, entering our computation at \LO and \NLO, are presented in the second column of \cref{tab:channels}, while corresponding representative diagrams at tree- and loop-level are displayed in \cref{fig:diagH3j_gg}.

The above discussion can be implemented using an effective Lagrangian including both decay categories:
\begin{equation}
  \cL_\mathrm{Higgs} = -\frac{\lambda(M_\Pqt,\muR)}{4}HG^{a}_{\mu\nu}G^{a,\mu\nu} + \frac{\yb(\muR)}{\sqrt{2}}H\bar\psi_\Pqb\psi_\Pqb \, .
  \label{eq:effLagrangian}
\end{equation}
Here, the effective Higgs-gluon coupling proportional to $\alphas$ is written in terms of the Higgs vacuum expectation value $v$ as,
\begin{equation}
  \lambda(M_\Pqt, \muR) = -\frac{\alphas(\muR)C(M_\Pqt,\muR)}{3\uppi v}
\end{equation}
and the $\PH\Pqb\Paqb$ Yukawa coupling is given as
\begin{equation}
  \yb(\muR) = \bar{m}_\Pqb(\muR)\frac{4\uppi\alpha}{\sqrt{2}M_\PW \sin\theta_\mathrm{W}} \, .
\end{equation}
As indicated by the $\muR$ arguments, these quantities are subject to renormalisation.
Throughout, we work in the \MSbar scheme and evaluate renormalised quantities at the renormalisation scale $\muR$ with $\NF = 5$.
While the top-quark Wilson coefficient $C(M_\Pqt,\muR)$ is known up to order $\alphas^4$ \cite{Inami:1982xt,Djouadi:1991tk,Chetyrkin:1997iv,Chetyrkin:1997un,Chetyrkin:2005ia,Schroder:2005hy,Baikov:2016tgj}, it suffices to consider the first-order expansion for our purposes, \ie,
\begin{equation}
  C(M_\Pqt,\muR) = 1 + C^{(1)}(M_\Pqt,\muR)\frac{\alphas(\muR)}{2\uppi} + \cO(\alphas^2) = 1 + \frac{11}{6}\NC\frac{\alphas(\muR)}{2\uppi} + \cO(\alphas^2) \, ,
  \label{eq:wilsonCoeff}
\end{equation}
which is independent of the top-quark mass $M_\Pqt$.
The running of the Yukawa mass $\bar{m}_\Pqb$ is taken into account using the results of \cite{Vermaseren:1997fq}.

Before further discussing the details of our calculation for each decay category, we wish to highlight that the two terms present in \cref{eq:effLagrangian} do not interfere in the approximation of kinematically massless quarks, as considered here. When computing squared amplitudes, the operators present in \cref{eq:effLagrangian} neither interfere nor mix with each other under renormalisation \cite{Gao:2019mlt}. 
In particular, this allows us to calculate higher-order QCD corrections for both decay channels independently and it renders the separate treatment of parton-level processes well-defined at each order.

We further wish to mention that in order to compare results in the $\PH\to\Pqb\Paqb$ and $\PH\to\Pg\Pg$ decay modes, we will rescale the differential decay rates computed in both categories by their respective branching ratios, defined as
\begin{align}
  \BrHbb^n(s,\muR) &= \frac{\GammaHbb^n(s,\muR)}{\GammaHbb^n(s,\muR)+\GammaHgg^n(s,\muR)} \, , \label{eq:brHbb}\\
  \BrHgg^n(s,\muR) &= \frac{\GammaHgg^n(s,\muR)}{\GammaHbb^n(s,\muR)+\GammaHgg^n(s,\muR)} \, , \label{eq:brHgg}
\end{align}
with $n=0,1$ to be considered at \LO and \NLO, respectively. 
At \LO, the inclusive hadronic $\PH\to 2j$ decay widths read
\begin{equation}
 \GammaHbb^0(\muR) = \frac{y_b^2(\muR)M_\PH\NC}{8\uppi} \, , \quad \GammaHgg^0(\muR) = \frac{\alphas^2(\muR)\GF M_\PH^3}{36\uppi^3\sqrt{2}} \, ,
\end{equation}
where $\GF$ is the Fermi constant and $M_\PH$ the Higgs mass. Their first-order QCD corrections are known since long, \cf \eg \cite{Gorishnii:1990zu,Gorishnii:1991zr,Kataev:1993be,Surguladze:1994gc,Larin:1995sq,Chetyrkin:1995pd,Chetyrkin:1996sr,Baikov:2005rw,Spira:1995rr,Inami:1982xt,Chetyrkin:1997iv,Baikov:2006ch}, and are given by
\begin{align}
 \GammaHbb^1(s,\muR) &= \GammaHbb^0(\muR)\left[1 + \left(\frac{\alphas(\muR)}{2\uppi}\right)\left(\frac{17}{2}\CF + 3\CF\log\left(\frac{\muR^2}{s}\right) \right) \right] \, \\
 \GammaHgg^1(s,\muR) &= \GammaHgg^0(\muR)\left[1 + \left(\frac{\alphas(\muR)}{2\uppi}\right)\left(\frac{95}{6}\CA - \frac{7}{3}\NF + 2\beta_0\log\left(\frac{\muR^2}{s}\right) \right) \right] \, .
\end{align}
The rescaling factors given in \cref{eq:brHbb,eq:brHgg} will be applied in \cref{sec:results}, where we present event-shape predictions for both decay modes.

To obtain our predictions for the Higgs event-shape distributions, we implement our \NLO QCD calculation in the publicly available\footnote{\url{http://eerad3.hepforge.org}} \eerad program \cite{Gehrmann-DeRidder:2014hxk}. This code has previously been used to study event shapes \cite{Gehrmann-DeRidder:2007vsv,Gehrmann-DeRidder:2009fgd} and jet distributions \cite{Gehrmann-DeRidder:2008qsl} in $\Pep\Pem \to 3j$ at \NNLO.
The implementation is done in a flexible manner, utilising the existing infrastructure for three-jet production in electron-positron annihilation and amending it by new subroutines for Higgs decays.
All matrix elements are implemented in analytic form, enabling a fast and numerically stable evaluation of the perturbative coefficients.
With this implementation, \eerad is promoted to a multi-process event generator, capable of calculating both off-shell $\PZ/\Pgg$ decays at \NNLO and Higgs decays at \NLO.

In the remainder of this section, we split the discussion of the ingredients of our computation in the following parts: 
in \cref{subsec:formalism}, we describe the general framework in which Higgs-decay observables are computed up to \NLO, while in \cref{subsec:Hbb,subsec:Hgg} we present all ingredients needed in the two Higgs decay categories.
We conclude in \cref{subsec:checks} by explicitly summarising all checks that we have performed to confirm the correctness of our results.

\subsection{General framework}
\label{subsec:formalism}
For an infrared-safe observable $O$, the differential decay rate of the Higgs boson to three-jet-like final states, 
normalised to the respective Born-level $\PH\to 2j$ decay width, can be written up to \NLO in the strong coupling $\alphas$ as,
\begin{equation}
  \frac{1}{\Gamma^n(s,\muR)}\frac{\rd \Gamma(s,\muR,O)}{\rd O} = \frac{\Gamma^0(\muR)}{\Gamma^n(s,\muR)}\left(\frac{\alphas(\muR)}{2\uppi}\right)\frac{\rd A(s)}{\rd O} + \frac{\Gamma^0(\muR)}{\Gamma^n(s,\muR)}\left(\frac{\alphas(\muR)}{2\uppi}\right)^2\frac{\rd B(s,\muR)}{\rd O} \, .
  \label{eq:rate}
\end{equation}
For \NLO results, the differential decay rate is normalised to $\Gamma^1$, obtained by taking $n=1$ in \cref{eq:rate}, while the additional ratios are unity for $n=0$, \ie, for \LO results.
Here, $A$ and $B$ denote the \LO and \NLO coefficients, respectively.

The \LO coefficient $A$ is determined by
\begin{equation}
  \left(\frac{\alphas(\muR)}{2\uppi}\right)\frac{\rd A(s)}{\rd O} =\frac{1}{\Gamma^0(\muR)}\int \frac{\rd \Gamma^{\mathrm{B}}(s)}{\rd \Phi_3}\delta\left(O - O(\Phi_3)\right)\, \rd\Phi_3 \, ,
  \label{eq:coeffLO}
\end{equation}
where $\frac{\rd\Gamma^\mathrm{B}(s)}{\rd\Phi_3}$ denotes the tree-level three-parton decay rate differential in the three-particle phase space.
As long as $\rd\Gamma^\mathrm{B}$ contains a suitable observable definition $\cJ{3}{3}(p_1,p_2,p_3)$, which ensures that all three final-state partons are well-enough separated, the \LO coefficient $A$ is infrared-finite.

The \NLO coefficient $B$ is calculated as
\begin{multline}
  \left(\frac{\alphas(\muR)}{2\uppi}\right)^2\frac{\rd B(s,\muR)}{\rd O} = \frac{1}{\Gamma^0(\muR)}\int \left[\frac{\rd\Gamma^\mathrm{V}(s,\muR)}{\rd \Phi_3} + \int \frac{\rd\Gamma^\mathrm{T}_{\NLO}(s,\muR)}{\rd\Phi_3} \right] \delta\left(O-O(\Phi_3)\right)\, \rd\Phi_3 \\
  + \frac{1}{\Gamma^0(\muR)}\int \left[\frac{\rd \Gamma^{\mathrm{R}}(s)}{\rd \Phi_4}\delta\left(O - O(\Phi_4)\right) - \frac{\rd \Gamma^{\mathrm{S}}_{\NLO}(s)}{\rd \Phi_4}\delta\left(O - O(\Phi_3(\Phi_4))\right)\right] \, \rd\Phi_4 \, ,
  \label{eq:coeffNLO}
\end{multline}
where the real and virtual subtraction terms $\rd\Gamma^\mathrm{S}_{\NLO}$ and $\rd\Gamma^\mathrm{T}_{\NLO}$ ensure that the real and virtual (one-loop) contributions denoted by $\mathrm{R}$ and $\mathrm{V}$ are separately infrared finite.
While the real subtraction term cancels singularities in the real contribution when one parton becomes unresolved, the virtual subtraction term cancels explicit infrared singularities arising from one-loop integrals in the virtual contribution.
We perform the subtraction in \cref{eq:coeffNLO} using the antenna-subtraction formalism \cite{Campbell:1998nn,Gehrmann-DeRidder:2005btv,Gehrmann-DeRidder:2007foh}.
Details on the construction of both types of subtraction terms can be found in \cite{Gehrmann-DeRidder:2005btv}.

While in principle all ingredients to our calculation are known for quite some time already, they have so far not been combined to obtain predictions for the observables considered here.

\begin{table}[t]
  \centering
  \caption{Partonic channels contributing to the decay $\PH \to 3j$ at \LO and \NLO.}
  \begin{tabular}{llll}\toprule
    ~ & $\PH\to\Pqb\Paqb$ type & $\PH\to\Pg\Pg$ type & ~\\ \midrule
    \LO & $\PH \to \Pqb \Paqb \Pg$ & $\PH \to \Pg \Pg \Pg$ & tree-level \\
    ~ & ~ & $\PH \to \Pg \Pq \Paq$ & tree-level \\ \midrule
    \NLO & $\PH \to \Pqb \Pqb \Pg$ & $\PH \to \Pg \Pg \Pg$ & one-loop \\
    ~ & ~ & $\PH \to \Pg \Pq \Paq$ & one-loop \\
    ~ & $\PH \to \Pqb \Paqb \Pg \Pg$ & $\PH \to \Pg \Pg \Pg \Pg$ & tree-level \\
    ~ & $\PH \to \Pqb \Paqb \Pq \Paq$ & $\PH \to \Pg \Pg \Pq \Paq$ & tree-level \\
    ~ & $\PH \to \Pqb \Paqb \Pqb \Paqb$ & $\PH \to \Pq \Paq \Pq' \Paq'$ & tree-level \\ \bottomrule
  \end{tabular}
  \label{tab:channels}
\end{table}

\subsection{Yukawa-type contributions}
\label{subsec:Hbb}
Representative Feynman diagrams of all partonic contributions in the $\PH\to\Pqb\Paqb$ channel are shown in \cref{fig:diagH3j_bb}, with Born-level and virtual contributions displayed in the first line and real contributions in the second line.
We calculate tree-level matrix elements with up to four partons explicitly using \form \cite{Kuipers:2012rf}, taking into account all processes given in \cref{tab:channels}.
The three-parton one-loop matrix element is adapted from the real-virtual corrections of the Higgs-decay part in the \NNLO $\PV\PH$ calculation implemented in \nnlojet \cite{Gauld:2019yng}, which implements the results of \cite{Anastasiou:2011qx,DelDuca:2015zqa}.
The real \NLO subtraction term has first been constructed as the single-unresolved part of the double-real subtraction term in the Higgs-decay part in \cite{Gauld:2019yng}, while the virtual \NLO subtraction term corresponds to the explicit-pole part of the real-virtual subtraction contribution in that calculation.

\subsection{Effective-theory-type contributions}
\label{subsec:Hgg}
Representative Feynman diagrams of all partonic contributions in the $\PH\to\Pg\Pg$ category are shown in \cref{fig:diagH3j_gg}, with Born-level and virtual contributions shown in the first two lines and real contributions in the third line. 
We construct tree-level matrix elements with up to four partons directly from gluon-gluon antenna functions \cite{Gehrmann-DeRidder:2005alt} taking into account the processes given in \cref{tab:channels}.
The HEFT three-parton one-loop matrix elements are adapted from the \mcfm $\PH+1j$ implementation \cite{Boughezal:2016wmq,Campbell:2019dru}, in turn based on the results given in \cite{Schmidt:1997wr}.
We note further that the $\cO(\alphas^2)$ correction to $\PH\to 3j$ present in the $\PH\to\Pg\Pg$ category at \NLO receives a contribution from the $\cO(\alphas)$ expansion of the Wilson coefficient $C(M_\Pqt,\muR)$ given in \cref{eq:wilsonCoeff}.
We implement this as a finite shift in the virtual correction
\begin{equation}
  \rd\Gamma^\mathrm{V}(s,\muR) \to \rd\Gamma^\mathrm{V}(s,\muR) + 2C^{(1)}(M_\Pqt,\muR)\frac{\alphas(\muR)}{2\uppi}\rd\Gamma^\mathrm{B}(s) \, .
\end{equation}

Taking into account the crossing of the amplitudes, all HEFT one-loop amplitudes are in agreement with the respective Higgs-production one-loop matrix elements available in \nnlojet \cite{Chen:2016zka}.
The real and virtual \NLO subtraction terms for gluonic Higgs decays have first been derived in \cite{Coloretti2021msc}, based on analytic expressions for the definition of the gluon-gluon antenna functions from Higgs boson decay contained  in \cite{Gehrmann-DeRidder:2005alt}.

\subsection{Validation}
\label{subsec:checks}
To conclude this section, we elaborate on the analytical and numerical checks we performed.

We validated the correctness of the implementation of all matrix elements numerically on a point-by-point basis. To this end, all tree-level matrix elements have been compared against \madgraph \cite{Alwall:2011uj,Alwall:2014hca}, the HEFT Higgs-decay one-loop matrix elements against \openloops2 \cite{Buccioni:2019sur}, and the Yukawa-induced one-loop matrix elements against \powhegbox \cite{Bizon:2019tfo}. In all cases, we have found perfect agreement with the reference results up to machine precision.

Our implementation of the relevant differential \NLO antenna-subtraction terms $\rd\Gamma^\mathrm{S}_{\NLO}$ has been checked numerically on a point-by-point basis against the real correction $\rd\Gamma^\mathrm{R}$ . For this purpose, we have used so-called ``spike-test'' checks, as defined in the context of the antenna-subtraction framework for the first time in \cite{Pires:2010jv} and applied in the context of the \NNLO di-jet computation in \cite{NigelGlover:2010kwr}.
In all cases, we found excellent agreement between the subtraction terms and real-radiation matrix elements in the relevant single-unresolved limits.
Finally, to confirm the analytic cancellation of explicit poles between virtual subtraction terms $\rd\Gamma^\mathrm{T}_{\NLO}$ and virtual corrections $\rd\Gamma^\mathrm{V}$, we have implemented numerical tests of the cancellation that are performed during actual runs.

\section{Numerical Results}
\label{sec:results}
We here focus on the classical six event shape observables related to hadronic Higgs decays. 
Those have been studied extensively in $\Pep\Pem$ collisions at LEP \cite{ALEPH:1996sio,ALEPH:2003obs,Wicke:1999bdp,DELPHI:2003yqh,DELPHI:2004omy,L3:1995eyy,L3:1997bxr,L3:1998ubl,L3:2002oql,L3:2004cdh,OPAL:1993pnw,OPAL:1996fae,OPAL:1997asf,OPAL:1999ldr,OPAL:2004wof}. 

After presenting the numerical set-up in \cref{subsec:setup}, we shall divide the discussion of our results for the 
six major event-shape observables into two subsections: \cref{subsec:Higgs_only}, where only the two Higgs decay 
categories are considered and \cref{subsec:comparisonZ}, where these predictions are also compared with those obtained for hadronic $\PZ/\Pgg$ decays.

\subsection{Numerical set-up and scale-variation prescription} 
\label{subsec:setup} 
We consider Higgs decays with a centre-of-mass energy $\sqrt{s} = M_\PH$, equal to a Higgs mass of $M_\PH = 125~\GeV$. Electroweak quantities are considered as constant parameters: specifically, the electromagnetic coupling $\alpha$ is taken to $\alpha = \frac{1}{128}$ and the weak-boson masses to
\begin{equation}
  M_\PZ = 91.1876~\GeV \, , \quad M_\PW = 80.385~\GeV \, .
\end{equation}

For the QCD part, all quantities are renormalisation-scale dependent and a scale-variation prescription is applied.
The renormalisation scale $\muR$ is varied as a multiple $\muR \to k_\mu \muR$ with $k_\mu \in \left[\frac{1}{2},2\right]$ of the on-shell Higgs-boson mass $M_\PH$, which we choose as our central scale, $\muR = M_\PH$.
We use a one and two-loop running for the strong coupling $\alphas$, at \LO and \NLO respectively, with a nominal value at scale $M_\PZ$ given by $\alphas(M_\PZ) = 0.1189$. The former is obtained by solving the renormalisation-group equation at the given order, as detailed in \cite{Gehrmann-DeRidder:2007vsv,Gehrmann-DeRidder:2014hxk}.

Further, throughout the computation, we keep a vanishing kinematical mass of the $\Pqb$-quark, but consider a non-vanishing Yukawa mass. We calculate the latter using the result of \cite{Vermaseren:1997fq}, corresponding to $\bar{m}_\Pqb(M_\PH)$ close to $2.61~\GeV$.

Fixed-order calculations of event shapes are generically only valid away from the two-jet limit, as they obtain large logarithmic corrections of the form $\alphas^mL^n$, where $L \sim \log(1/O)$, when additional particles become unresolved \cite{Catani:1992ua}.
In order to obtain reliable predictions in the two-jet region, these logarithms have to be resummed, \cf \eg \cite{Monni:2011gb,Banfi:2014sua,Banfi:2016zlc,Baberuxki:2019ifp} for the combination of fixed-order and resummed predictions for event shapes in $\Pep\Pem$ annihilation.
As a consequence, we therefore impose a small cut-off $y_0 = 10^{-5}$ on all linear distributions and $y_0 = \re^{-10}$ on logarithmically binned distributions in practice.

\subsection{Differential predictions for Higgs-decay event-shape observables} 
\label{subsec:Higgs_only}
\begin{figure}[t]
  \centering
  \includegraphics[width=0.49\textwidth]{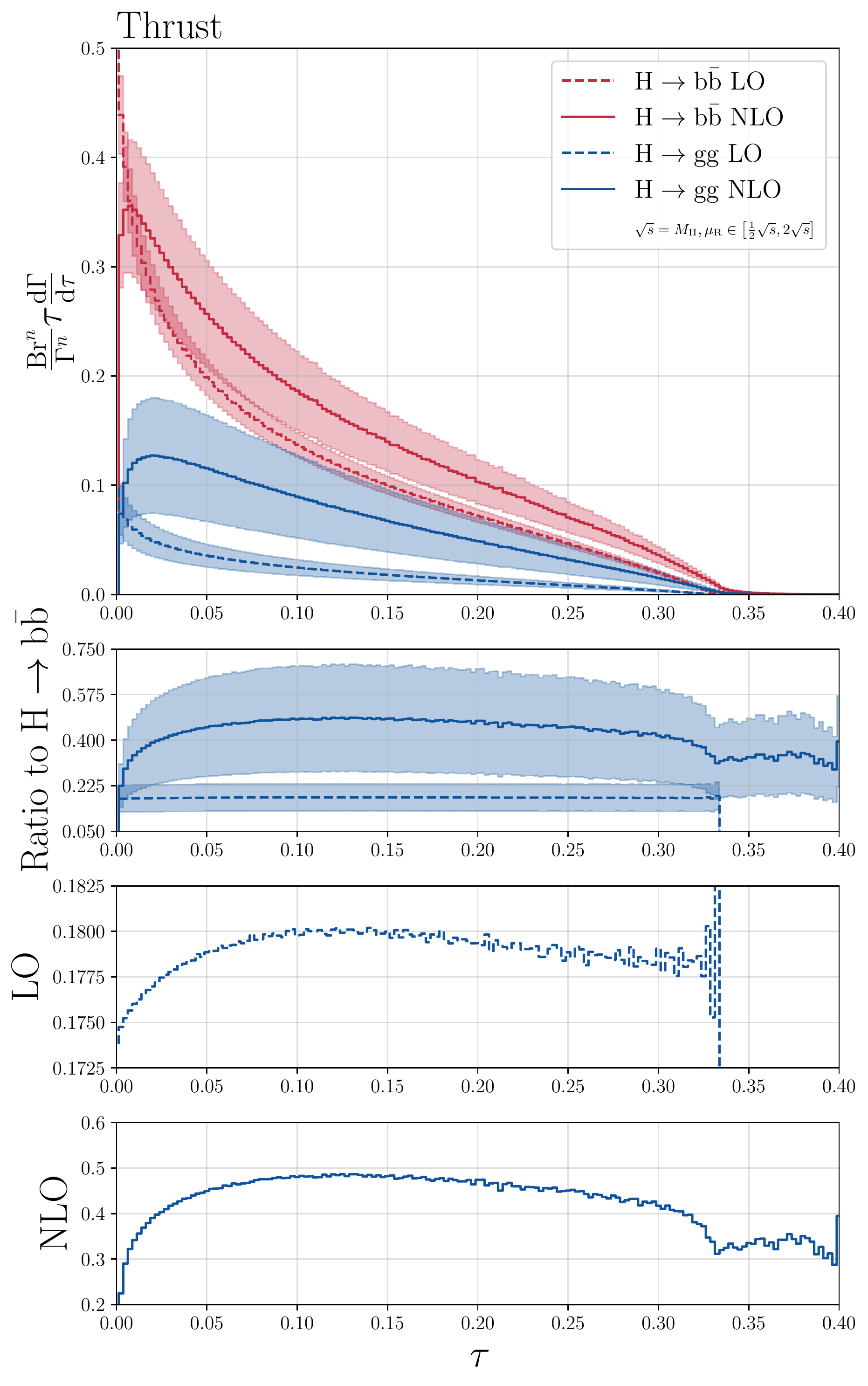}
  \includegraphics[width=0.49\textwidth]{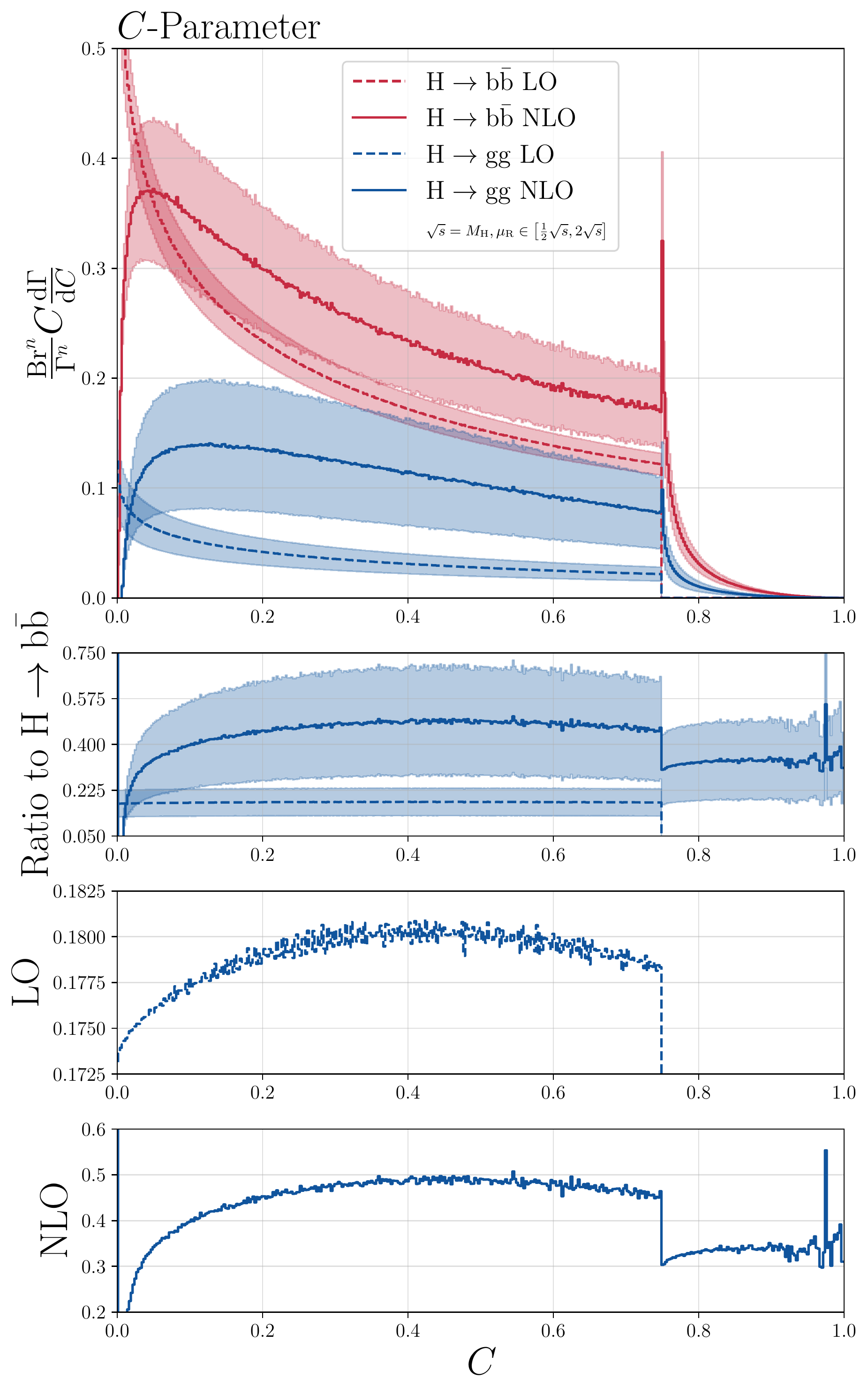}
  \caption{Thrust (\textit{left}) and $C$-parameter (\textit{right}). The upper panel contains predictions in the $\PH\to\Pg\Pg$ (\textit{blue}) and $\PH\to\Pqb\Paqb$ (\textit{red}) categories at \LO (\textit{dashed}) and \NLO (\textit{solid}), with renormalisation-scale uncertainties shown as light-shaded bands. The three lower panels show bin-by-bin ratios at \LO and \NLO, see main text.}
  \label{fig:thrustAndC}
\end{figure}

\begin{figure}[t]
  \centering
  \includegraphics[width=0.49\textwidth]{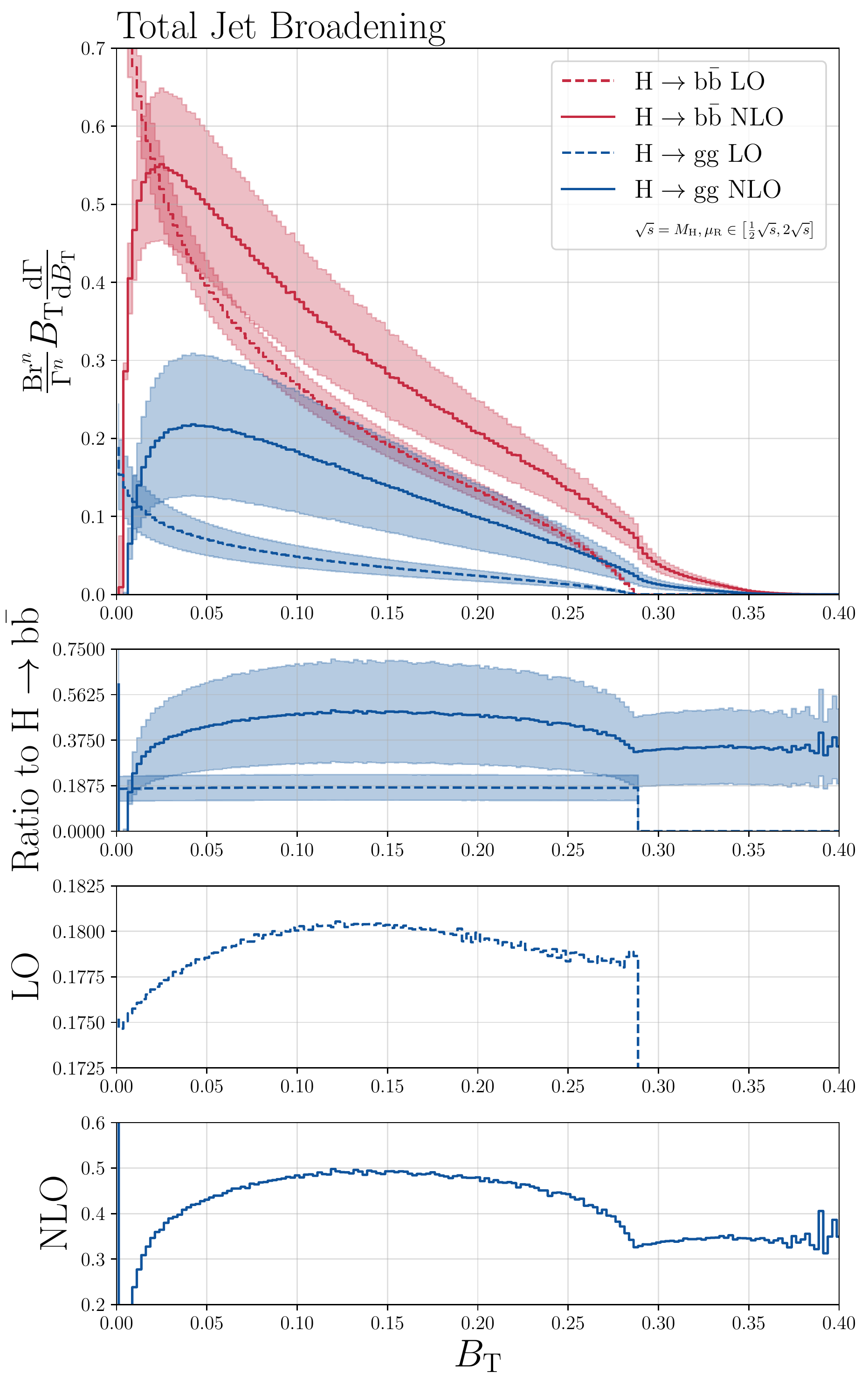}
  \includegraphics[width=0.49\textwidth]{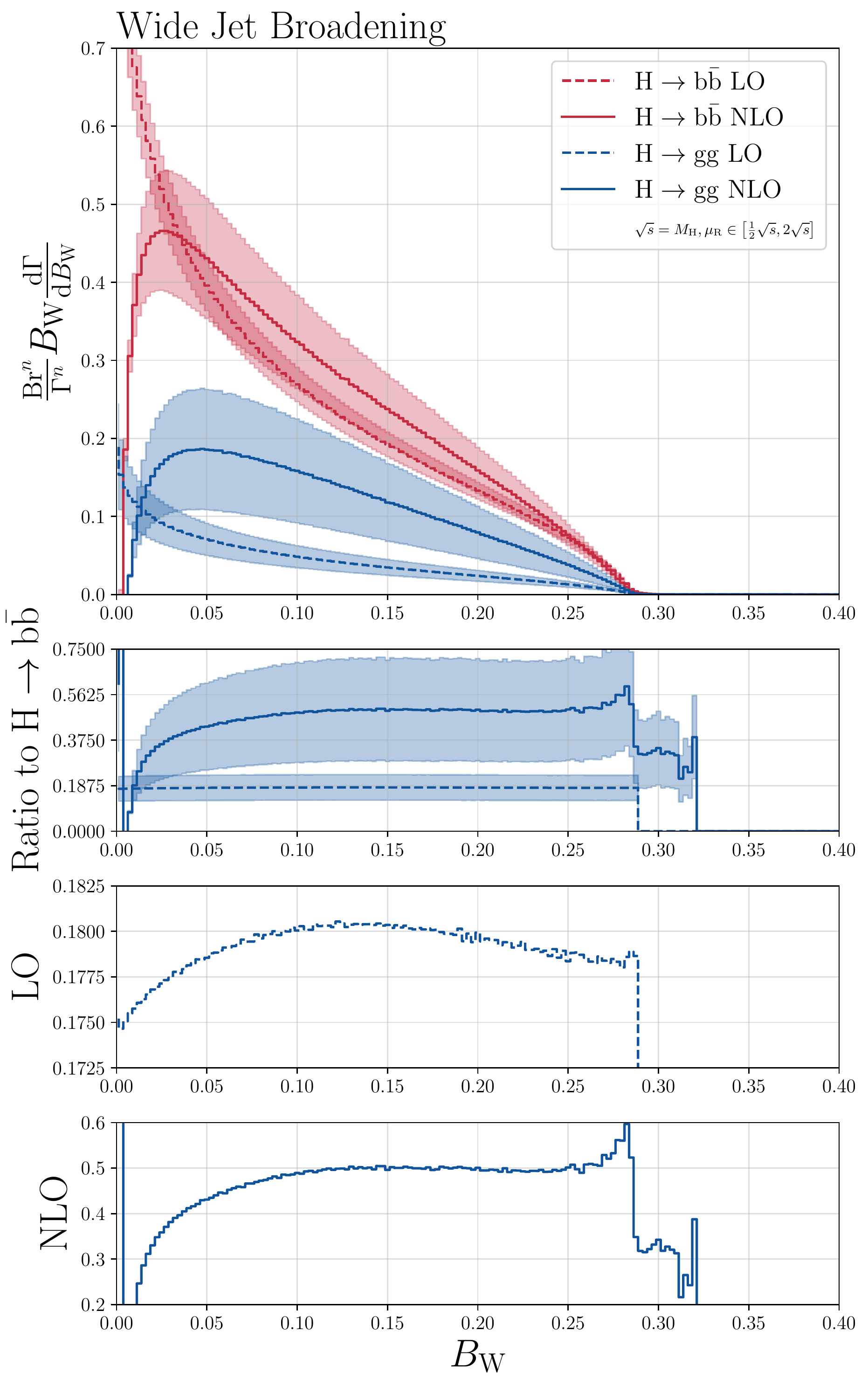}
  \caption{Total (\textit{left}) and wide jet broadening (\textit{right}). The upper panel contains predictions in the $\PH\to\Pg\Pg$ (\textit{blue}) and $\PH\to\Pqb\Paqb$ (\textit{red}) categories at \LO (\textit{dashed}) and \NLO (\textit{solid}), with renormalisation-scale uncertainties shown as light-shaded bands. The three lower panels show bin-by-bin ratios at \LO and \NLO, see main text.}
  \label{fig:broadenings}
\end{figure}

\begin{figure}[t]
  \centering
  \includegraphics[width=0.49\textwidth]{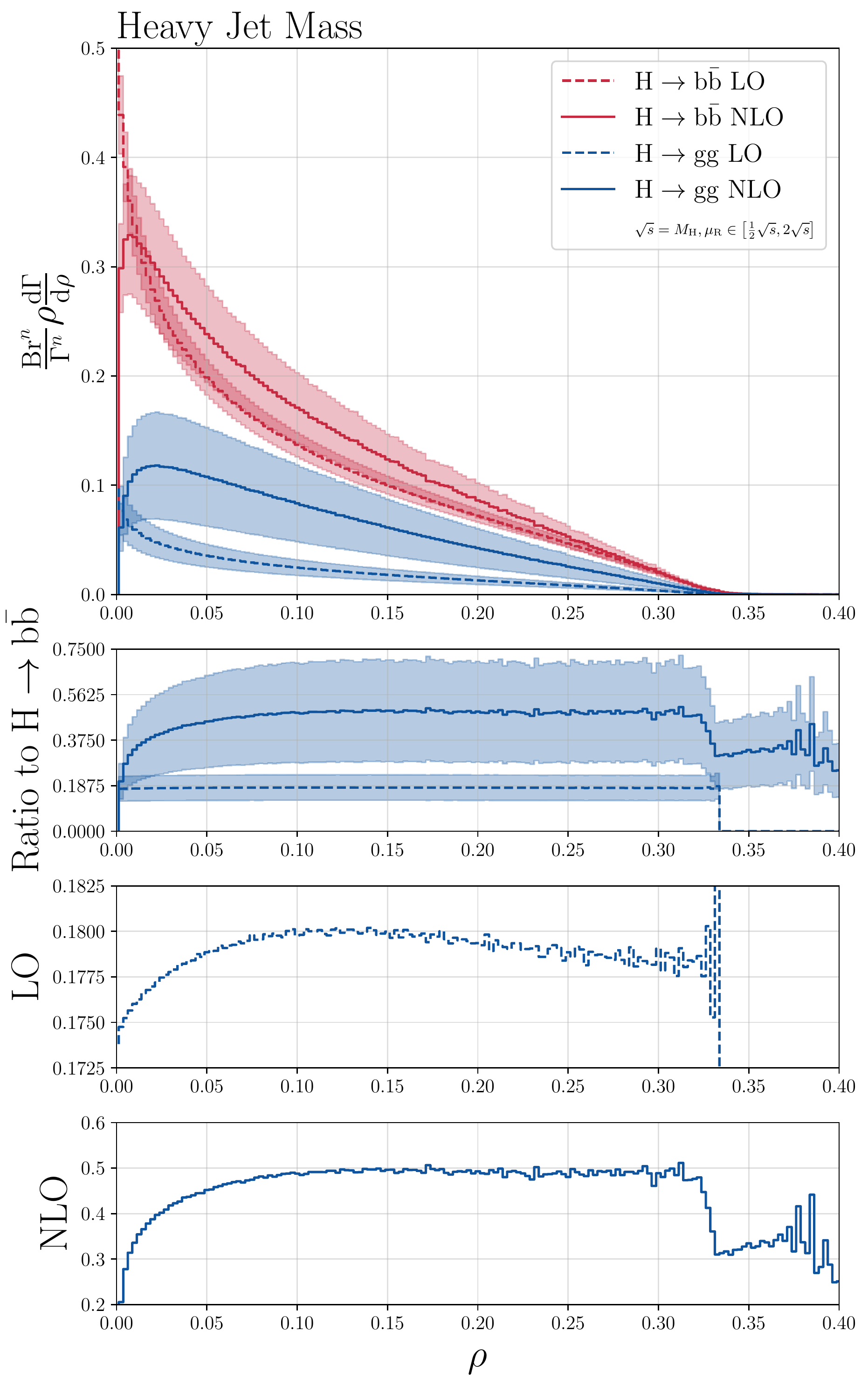}
  \includegraphics[width=0.49\textwidth]{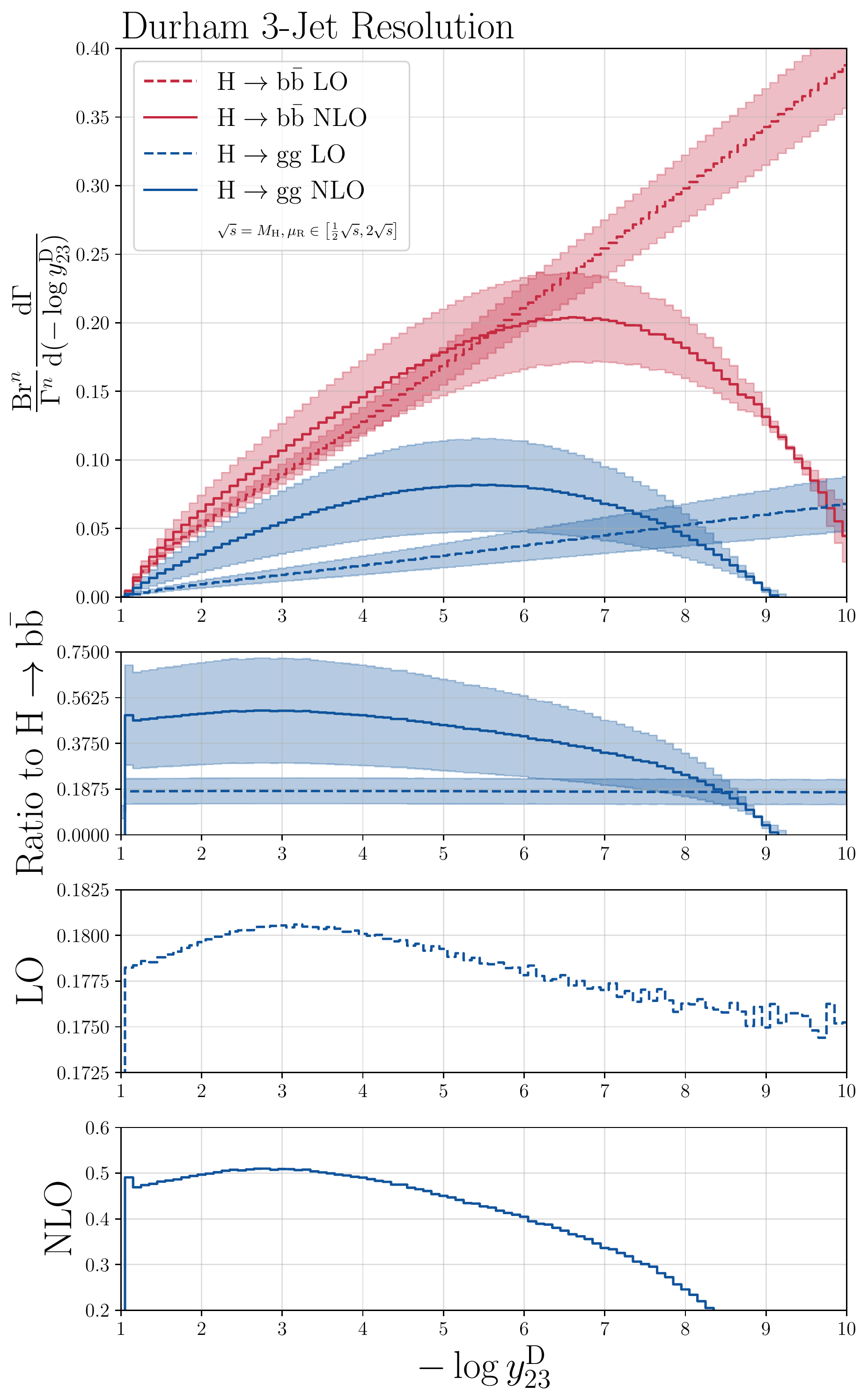}
  \caption{Heavy-jet mass (\textit{left}) and Durham three-jet resolution (\textit{right}). The upper panel contains predictions in the $\PH\to\Pg\Pg$ (\textit{blue}) and $\PH\to\Pqb\Paqb$ (\textit{red}) categories at \LO (\textit{dashed}) and \NLO (\textit{solid}), with renormalisation-scale uncertainties shown as light-shaded bands. The three lower panels show bin-by-bin ratios at \LO and \NLO, see main text.}
  \label{fig:rhoAndY}
\end{figure}

We here collect \LO and \NLO predictions for all six classical event shapes in \cref{fig:thrustAndC,fig:broadenings,fig:rhoAndY}, where we present observable-weighted distributions of the form,
\begin{equation}
  \frac{\Br^n(s,\muR)}{\Gamma^n(s,\muR)} O \frac{\rd\Gamma(s,\muR,O)}{\rd O} \, . \label{eq:dist}
\end{equation}
More specifically, according to \cref{eq:dist}, these normalised differential decay rates, given in \cref{eq:rate} are multiplied by the event-shape observable itself and scaled by the corresponding fixed-order branching ratio of the respective Higgs-decay category as given in \cref{eq:brHbb,eq:brHgg}:
\begin{align}
  & \BrHbb^0(s=M_\PH^2,M_\PH) \approx 0.933 \, , \quad \BrHgg^0(s=M_\PH^2,M_\PH) \approx 0.067 \, , \label{eq:Br0}\\
  & \BrHbb^1(s=M_\PH^2,M_\PH) \approx 0.908 \, , \quad \BrHgg^1(s=M_\PH^2,M_\PH) \approx 0.092 \, . \label{eq:Br1}
\end{align}
We have checked that the relative size of the $\PH\to\Pg\Pg$ and $\PH\to\Pqb\Paqb$ branching ratios in \cref{eq:Br0,eq:Br1} is in line with the results of \cite{Baglio:2010ae}.

In general, each of the figures, numbered as \cref{fig:thrustAndC,fig:broadenings,fig:rhoAndY}, are composed of four panels: the upper panel contains the distributions, while the lower three panels contain three ratios evaluated using differential results 
from the $\PH\to\Pg\Pg$ decay category in the numerator results from the $\PH\to\Pqb\Paqb$ decay category in the denominator. The first ratio panel contains both \LO and \NLO results, while the second and third panel zoom in on the \LO and \NLO results, respectively.
The line colour and style in the upper panel is chosen as follows: $\PH\to\Pqb\Paqb$ is shown in red, $\PH\to\Pg\Pg$ in blue; \NLO results have solid lines, and \LO results have dashed lines. In the lower panels, all ratio histograms inherit the colour of the numerator. Those are therefore all shown in blue. 
Uncertainties arising from a variation of the renormalisation scale in the aforementioned envelope $\muR \in \left[\frac{1}{2}\sqrt{s}, 2\sqrt{s} \right]$ are shown as light-shaded bands around the central results.
For the sake of clarity, uncertainty bands are only shown in the upper panel and in the joint (second panel) showing both \LO and \NLO ratios.

As can be inferred from the relative size of the branching ratios given in \cref{eq:Br0,eq:Br1}, it is found that the $\PH\to\Pqb\Paqb$ decay part is the dominant one in all distributions.
We observe sizeable \NLO corrections in both the di-gluon and the $\Pqb$-quark decay channel, with generally significantly larger corrections in the former.
In all cases, renormalisation-scale uncertainties are comparatively large at \NLO, indicating that \NNLO corrections can be expected to be non-negligible. 
Our findings regarding the large size of \NLO corrections in the di-gluon channel is not unexpected.
A similar behaviour is observed in Higgs production in gluon-gluon fusion, where at least \NNLO corrections are mandatory to obtain precise predictions, as first calculated in \cite{Anastasiou:2005qj}.
All distributions develop the expected perturbative properties that the differential cross sections diverge in the two-jet limit towards positive infinity at \LO, while at \NLO, they turn towards negative infinity. As a consequence, all distributions show the characteristic extremum towards the infrared region.

For all event-shape distributions considered here, the differential ratio of the $\PH\to\Pg\Pg$ to the $\PH\to\Pqb\Paqb$ result varies around $17.5\%-18.0\%$ at \LO and around $30\%-50\%$ at \NLO.
A very distinctive feature is also that at \LO level, the ratio between the $\PH\to\Pg\Pg$ and $\PH\to\Pqb\Paqb$ channels always develops an insignificant, seemingly constant shape with a plateau towards the multi-jet limit. While the \LO ratio shapes can be highlighted on a zoomed-in scale, these are overwhelmed by the size of the \NLO corrections.
In particular, this implies that the discriminative power of event shapes is significantly reduced if no higher-order corrections are taken into account, either via fixed-order corrections or (parton-shower) resummation.
In all cases, the shape of the ratio distribution at \NLO resembles the \LO one, however, significantly amplified and with shape differences towards the multi-jet limit.
Below, we will therefore limit the discussion to the results associated to each of the six event shape observables at \NLO level.

\paragraph{Thrust}
For multi-jet final states in lepton collisions, the thrust is defined as \cite{Brandt:1964sa,Farhi:1977sg}
\begin{equation}
  T = \max\limits_{\vec{n}}\left(\frac{\sum\limits_i \mods{\vec{p}_i\cdot\vec{n}}}{\sum\limits_i \mods{\vec{p}_i}} \right) \, ,
\end{equation}
where the sum of three-momenta is maximised over the direction of $\vec{n}$. The unit vector $\vec{n}_T$ which maximises the expression on the right-hand side defines the thrust axis. For two-jet events, the thrust approaches unity, $T\to 1$, while for three-jet events $T \geq \frac{2}{3}$. In practice, we consider the observable $\tau = 1-T$, so that $\tau > 0$ measures the departure from a two-particle topology.

The thrust distributions are shown in the left-hand panel of \cref{fig:thrustAndC}. In the region $0.05 \leq \tau \leq 0.25$, we find differential \NLO $K$-factors varying in the range $1.29-1.53$ in the $\PH\to\Pqb\Paqb$ decay category and between $3.2$ and $3.84$ in the $\PH\to\Pg\Pg$ category.

Our results are in agreement with the \NLO results presented in \cite{Gao:2019mlt}. 
Furthermore, we have implemented the analytic expressions provided in \cite{Gao:2019mlt} 
for both Higgs decay categories and found perfect agreement with our corresponding numerical results.

Our \NLO predictions for the thrust distribution have the following features: 
as shown in the left-hand panel of \cref{fig:thrustAndC}, for the di-gluon channel, the peak of the distribution is slightly shifted towards the hard region on the right-hand side of the plot.
Moreover, the distribution is broader in the $\PH\to\Pg\Pg$ channel, as is visible from the ratio plots in the lower panes of the figure. Despite generally looking rather flat above $\tau \approx 0.05$, the ratio to the $\Pqb\Paqb$ channel reaches a maximum at $\tau = 0.1-0.15$ and slowly falls off after.
While only developing a small shape difference in the intermediate $\tau$ region, the thrust distribution may therefore offer some potential for discriminating the Higgs-decay processes involving either heavy quarks or gluons, if this particular kinematical region is enhanced by suitably placed cuts.

\paragraph{$C$-Parameter}
The $C$-parameter,
\begin{equation}
  C = 3(\lambda_1\lambda_2 + \lambda_2\lambda_3 + \lambda_3\lambda_1) \, ,
\end{equation}
is defined in terms of the three eigenvalues $\lambda_{1,2,3}$ of the linearised momentum tensor \cite{Parisi:1978eg,Donoghue:1979vi}
\begin{equation}
  \Theta^{\alpha\beta} = \frac{1}{\sum\limits_j\mods{\vec{p}_j}}\sum\limits_j\frac{p_j^\alpha p_k^\beta}{\mods{\vec{p}_j}} \, , \text{where } \alpha,\beta \in \{1,2,3\} \, .
\end{equation}

Results for the $C$-parameter are presented in the right-hand pane of \cref{fig:thrustAndC}.
Clearly visible in both decay categories is the characteristic Sudakov shoulder \cite{Catani:1998sf} around $C\approx 0.75$, where large logarithms in the physical region spoil the convergence of fixed-order calculations.
In the region $0.1 \leq C \leq 0.6$, we find differential \NLO $K$-factors ranging from $1.17-1.40$ in the $\PH\to\Pqb\Paqb$ decay category and from $2.61-3.84$ in the $\PH\to\Pg\Pg$ category.

As for the thrust, the peak of the distribution is shifted towards the hard region on the right-hand side of the plot in the di-gluon channel.
It is to be noted, however, that the ratio plots differ considerably compared to the thrust case. The ratio to the $\Pqb$-quark channel distribution levels at values $C \gtrsim 0.2$, which limits the applicability of the $C$-parameter as a good observable for distinguishing the two Higgs-decay processes.

\paragraph{Jet Broadening} The total and wide jet broadening $B_\mathrm{T}$ and $B_\mathrm{W}$ are defined by \cite{Rakow:1981qn,Catani:1992jc}
\begin{equation}
  B_\mathrm{T} = B_1+B_2 \, \quad B_\mathrm{W} = \max(B_1,B_2) \, ,
\end{equation}
in terms of the hemisphere broadening
\begin{equation}
  B_i = \frac{\sum\limits_{j\in H_i}\mods{\vec{p}_i\times \vec{n}_T}}{2\sum\limits_j\mods{\vec{p}_j}} \, ,
\end{equation}
for two hemispheres $H_{1,2}$ separated by the thrust vector $\vec{n}_T$. Both vanish in the two-jet limit, $B_\mathrm{T}\to 0$, $B_\mathrm{W} \to 0$, and are bounded from above by $B_\mathrm{T} = B_\mathrm{W} = \frac{1}{2\sqrt{3}}$.

The jet-broadening distributions are shown in \cref{fig:broadenings}. Expectedly, both distributions have discontinuities at the three-jet maximum $\frac{1}{2\sqrt{3}} \approx 0.29$, where the \NLO calculation receives new contributions from four-particle configurations.
The \NLO corrections are generically smaller for the wide jet broadening than for the total jet broadening.
Specifically, for the total jet broadening, we find differential \NLO $K$-factors of around $1.25-1.55$ in the $\PH\to\Pqb\Paqb$ decay category and around $2.99-4.13$ in the $\PH\to\Pg\Pg$ category in the region $0.05 \leq B_\mathrm{T} \leq 0.2$. For the wide jet broadening, we find differential \NLO $K$-factors of around $1.06-1.24$ in the $\PH\to\Pqb\Paqb$ decay category and $2.58-3.44$ in the $\PH\to\Pg\Pg$ category in the region $0.05 \leq B_\mathrm{W} \leq 0.25$.

The ratio of the total-jet-broadening distributions develops a broad, however peaked, structure with a maximum around $B_\mathrm{T} \approx 0.15$, whereas the ratio in the case of the wide jet broadening distribution grows monotonically with a sharp peak at around $B_\mathrm{W}\approx 0.28$ and an inflection at around $B_\mathrm{W} \approx 0.15$. While the sharp peak in the wide-jet-broadening ratio plot may be considered useful for discrimination applications, it has to be noted that it lies in a region where higher-order corrections from fixed-order or resummed calculations have a large impact. In combination with the fact that this sharp peak is followed by a discontinuity (which vanishes upon resummation), we conclude that the wide jet broadening is not a suitable candidate for discrimination of Higgs decays to bottom quark pairs or gluons, because of the overall flat ratio between the two decay categories above $B_\mathrm{W} \approx 0.1$.
The total jet broadening, on the other hand, shows a similar shape difference between the two decay channels as the thrust distribution. As such, it may offer some potential to be used as a good discriminator if the region $0.05 \leq B_\mathrm{T} \leq 0.25$ can be suitably enhanced via well-placed cuts.

\paragraph{Heavy-jet mass}
In order to define the heavy-jet (or heavy-hemisphere) mass, the event is divided into two hemispheres $H_1$, $H_2$, for each of which the scaled invariant mass is calculated \cite{Clavelli:1981yh}:
\begin{equation}
  \frac{M_i^2}{s} = \frac{1}{E_\mathrm{vis}^2}\left(\sum\limits_{j \in H_i} p_j\right)^2 \, .
\end{equation}
We define the two hemispheres so that they are separated by a plane orthogonal to the thrust axis and calculate the heavy-jet mass as the maximum over the two invariant masses,
\begin{equation}
  \rho = \max_{i\in\{1,2\}}\left(\frac{M_i}{s}\right) \equiv \frac{M_\mathrm{H}^2}{s} \, .
\end{equation}
The $\rho$ and $\tau$ distributions are identical at \LO and as such vanish in the two-particle limit, $\rho \to 0$. For three-jet events, the heavy-jet mass is bounded by $\rho \leq \frac{1}{3}$.

The left-hand plot in \cref{fig:rhoAndY} contains the heavy-jet mass distributions in both Higgs-decay channels.
In the region $0.05 \leq \rho \leq 0.25$, we find comparatively modest differential \NLO $K$-factors of about $1.14-1.26$ in the $\PH\to\Pqb\Paqb$ decay category and $2.99-3.44$ in the $\PH\to\Pg\Pg$ category.

The ratio of the distributions obtained in the two categories remains almost exactly constant above $\rho\approx 0.1$, heavily limiting its application for disentangling the two Higgs-decay processes, as virtually no shape difference can be determined in the region $0.05 \leq \rho \leq 0.3$ in which the fixed-order calculation can be trusted.

\paragraph{3-Jet Resolution}
For a given jet algorithm, the three-jet resolution variable is given by that value for which an event is clustered from a three-jet event to a two-jet event.
In particular, we consider here the Durham jet algorithm, using the particle-distance measure \cite{Catani:1991hj,Brown:1990nm,Brown:1991hx,Stirling:1991ds,Bethke:1991wk}
\begin{equation}
  y_{ij}^{\mathrm{D}} = \frac{2\min (E_i^2,E_j^2)(1-\cos\theta_{ij})}{E_\mathrm{vis}^2} \, ,
\end{equation}
for each particle pair $(i,j)$. In the so-called ``E-scheme'', the pair with smallest resolution $y_{ij}^{\mathrm{D}}$ is replaced by a pseudo-particle with a four-momentum equal to the sum of the four-momenta of $i$ and $j$ at each step of the jet algorithm. As long as there are pairs with invariant mass below a cut $y_\mathrm{cut}$ in the event, this procedure is repeated.

The three-jet resolution scale in the Durham algorithm is shown in the right-hand part of \cref{fig:rhoAndY}.
In the region $2 \leq -\log(y_{23}^\mathrm{D}) \leq 5$, we find modest differential \NLO $K$-factors of about $1.07-1.21$ in the $\PH\to\Pqb\Paqb$ decay category and $2.71-3.39$ in the $\PH\to\Pg\Pg$ category.

In comparison to the previously discussed event-shape distributions, the three-jet clustering scale shows a slightly more complex shape difference between the di-gluon and quark-pair channel. While the peak of the $\PH\to\Pqb\Paqb$ distribution is located at around $\log y_{23}^\mathrm{D} \approx -6.5$, it is shifted to $\log y_{23}^\mathrm{D} \approx -5.5$ for the $\PH\to \Pg\Pg$ case. Moreover, the latter has a higher rate towards the hard region at the left-hand side of the plot, implying that gluon-induced jets are typically clustered earlier (at higher resolution scales) than quark-induced ones. This excess of gluon-induced jet clusterings is most pronounced at around $\log y_{23}^\mathrm{D} \approx -3$, as is visible from the first ratio panel in \cref{fig:rhoAndY}. In this phase-space region, the three-jet resolution parameter may therefore prove itself useful for disentangling Higgs decays related to bottom quark pairs and gluons.

\paragraph{Summary}
In conclusion, the event shapes studied here can be classified into two classes. We classify the observables depending on whether the ratio between the $\PH\to\Pg\Pg$ and $\PH\to\Pqb\Paqb$ result forms a plateau in the multi-jet limit or develops a maximum at intermediate event-shape values. While the former class offers virtually no potential for separating the contributions coming from Higgs decays to bottom quark pairs and gluons, the latter class is more suited towards this application due its richer structure.
Event shapes falling into the first class are the $C$-parameter, wide jet broadening, and the heavy-jet mass; event shapes of the second class are thrust, total jet broadening, and the Durham three-jet scale, \cf~\cref{fig:thrustAndC,fig:broadenings,fig:rhoAndY}.

\subsection{Comparison to $(\PZ/\Pgg)^*$ decays}
\label{subsec:comparisonZ}
\begin{figure}[t!]
  \centering
  \includegraphics[width=0.32\textwidth]{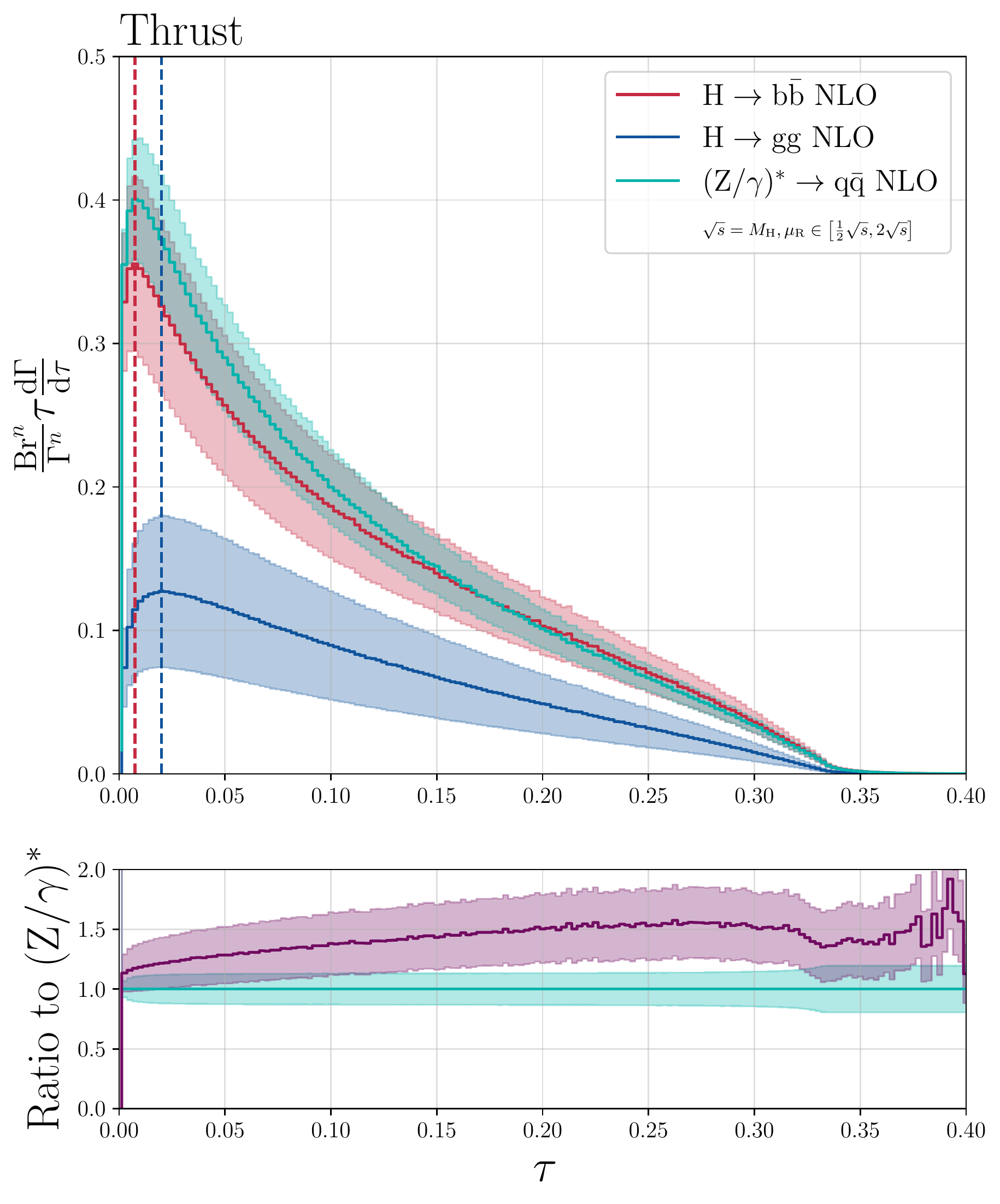}
  \includegraphics[width=0.32\textwidth]{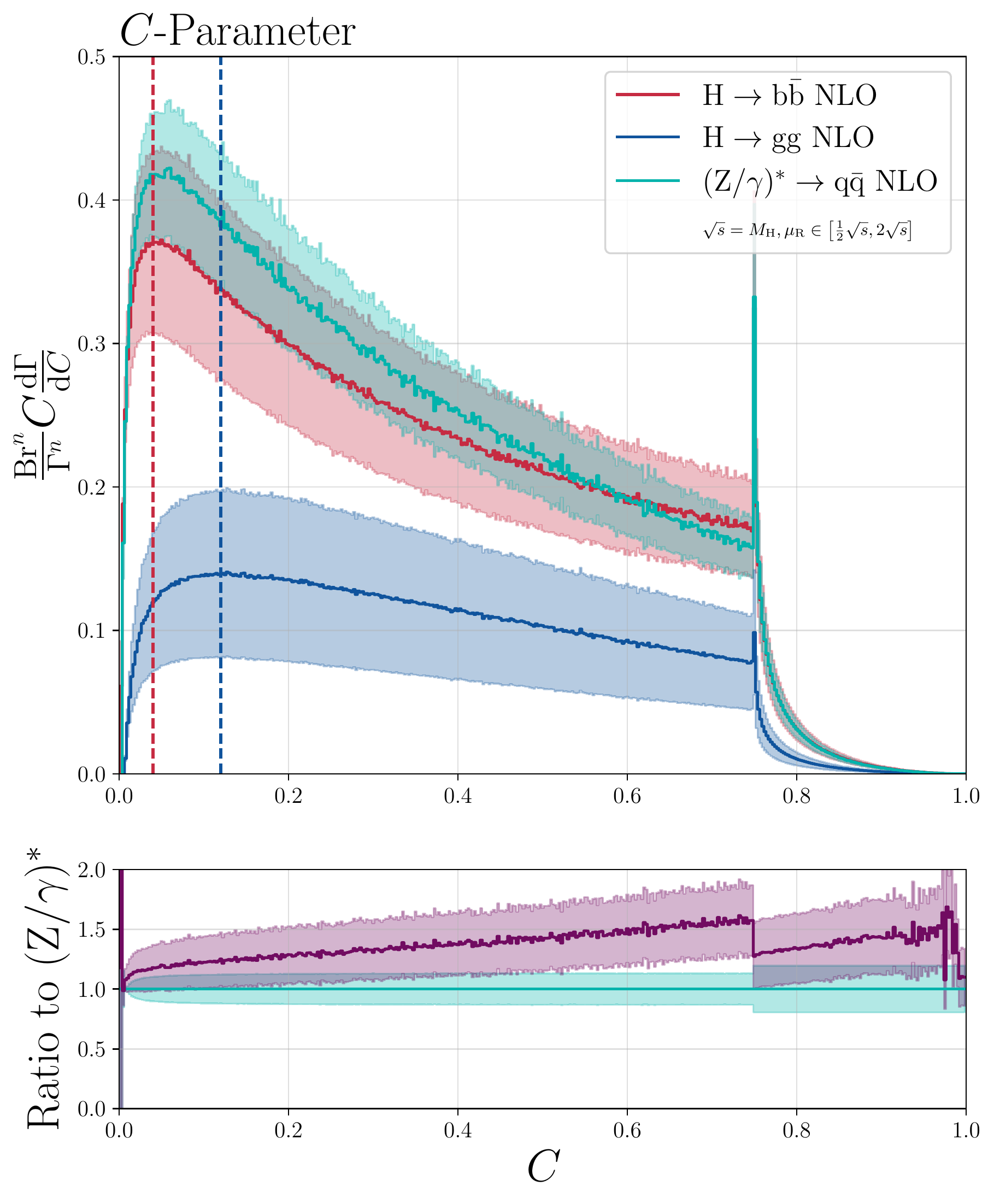}
  \includegraphics[width=0.32\textwidth]{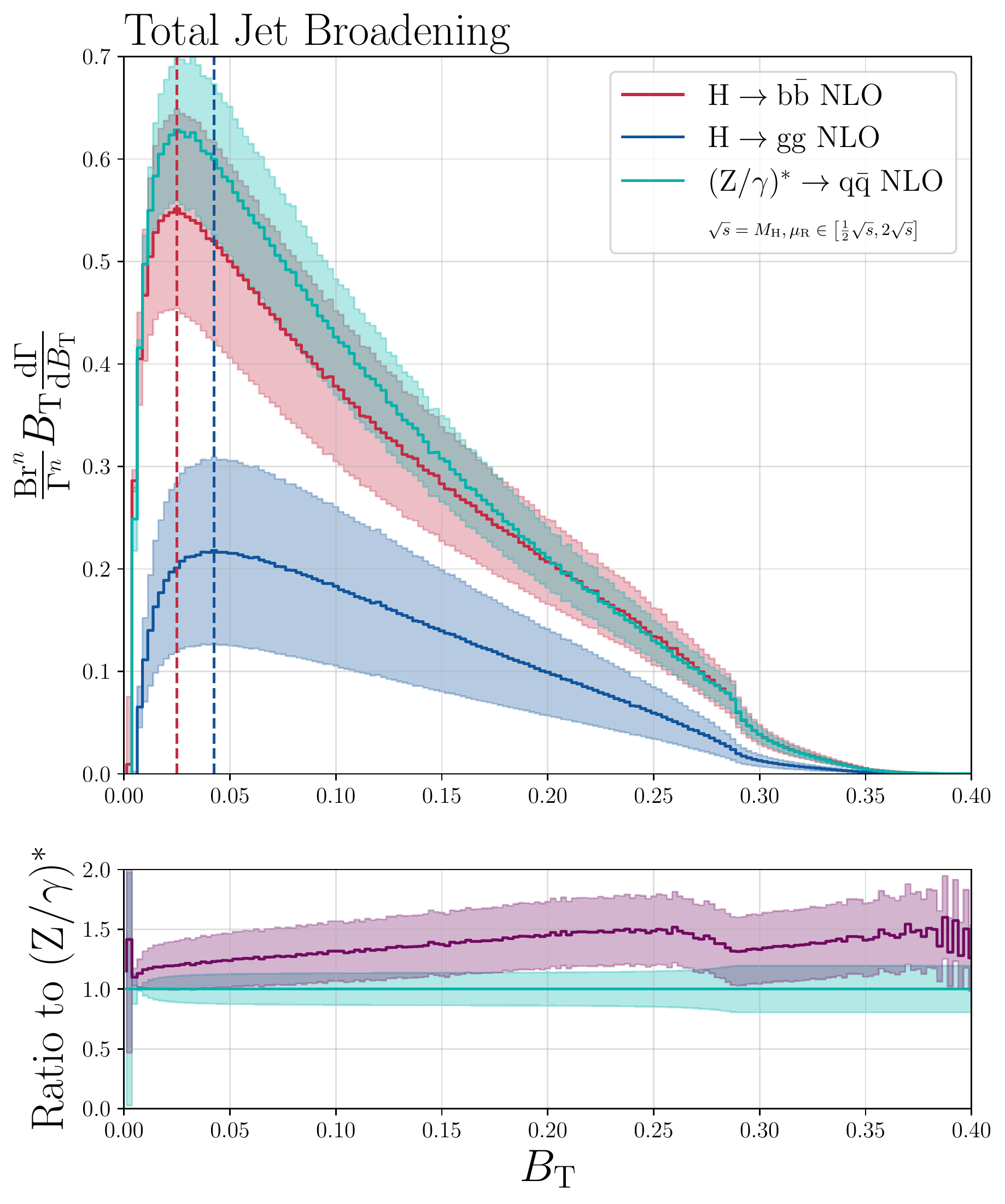}
  \includegraphics[width=0.32\textwidth]{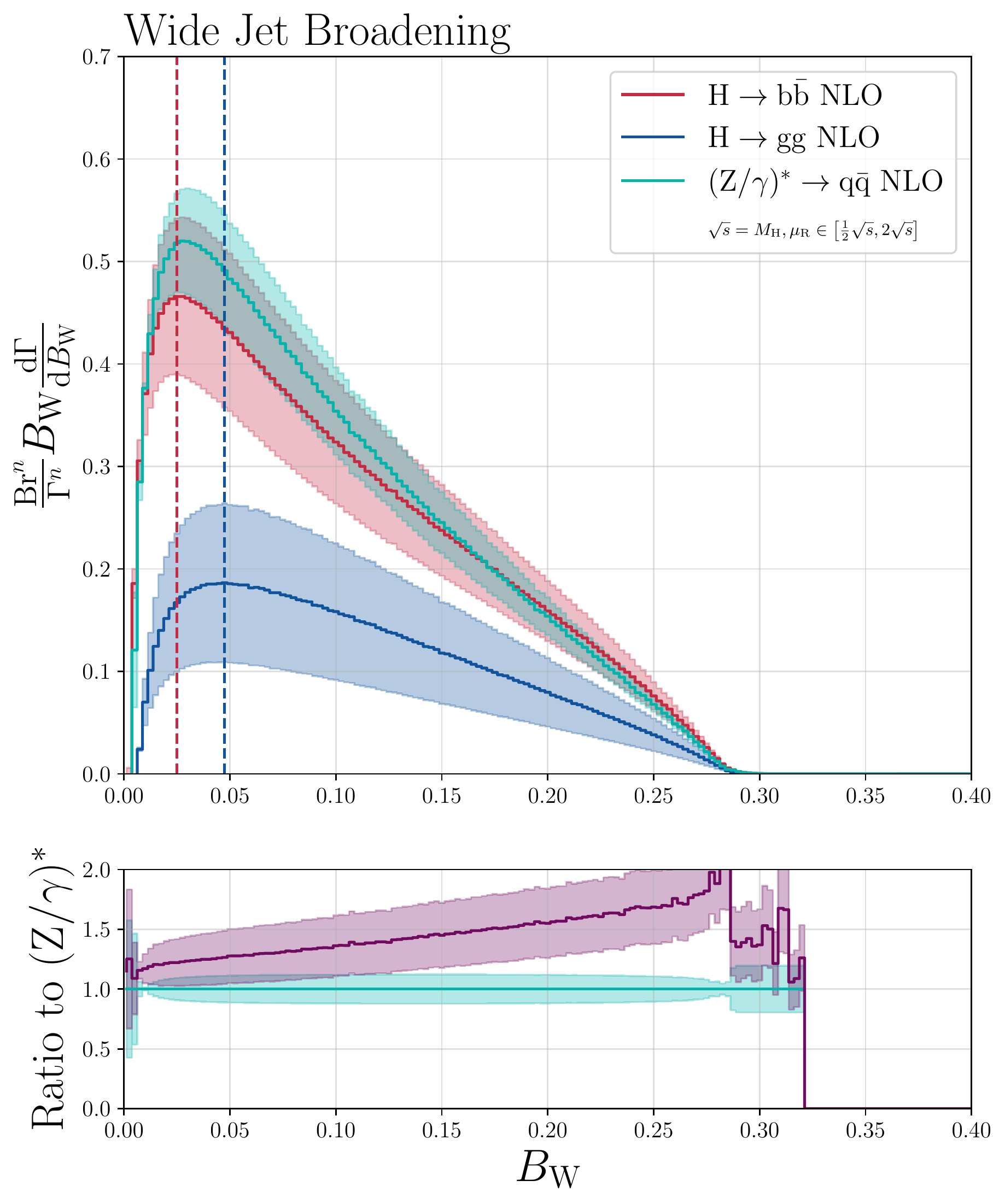}
  \includegraphics[width=0.32\textwidth]{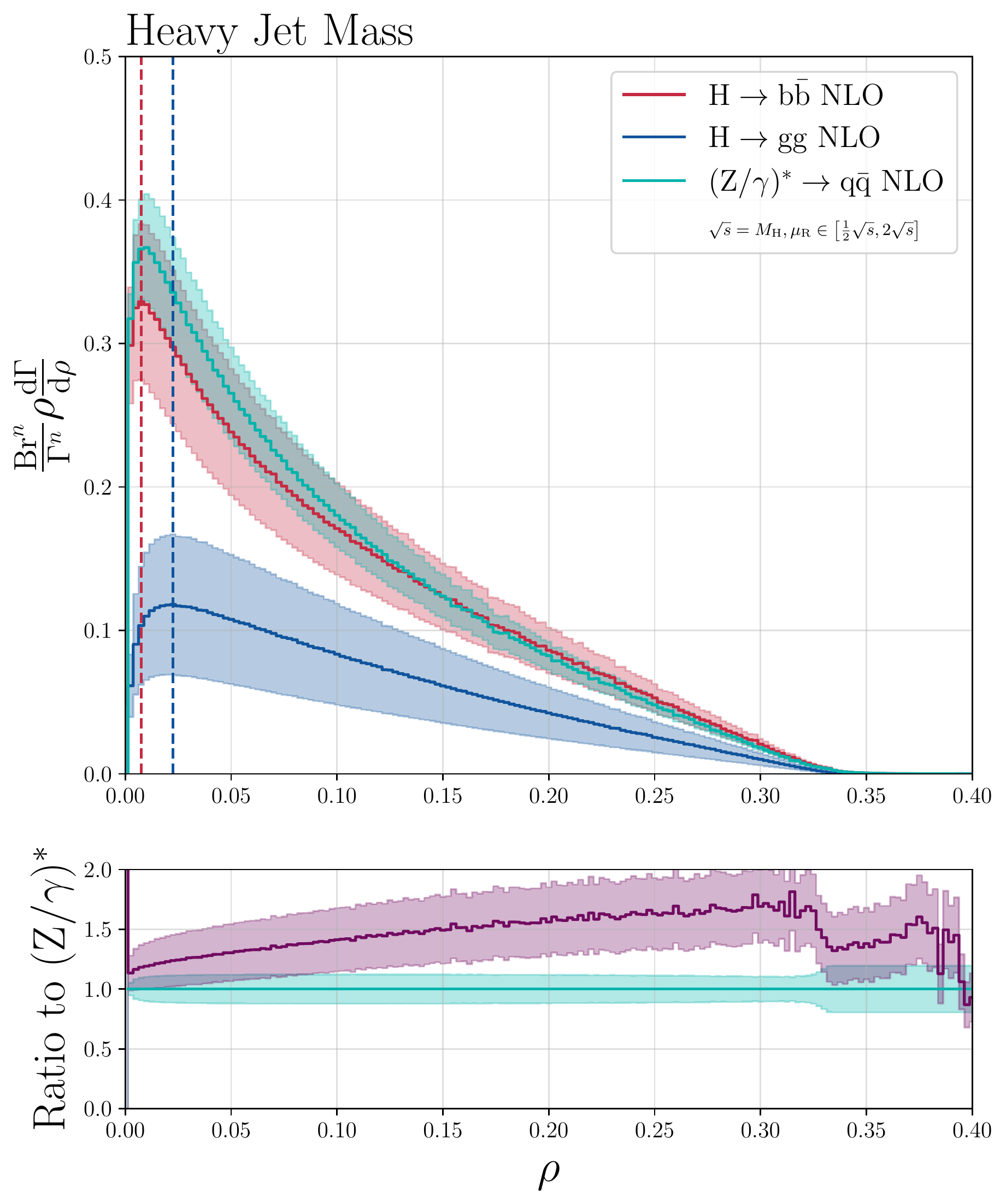}
  \includegraphics[width=0.32\textwidth]{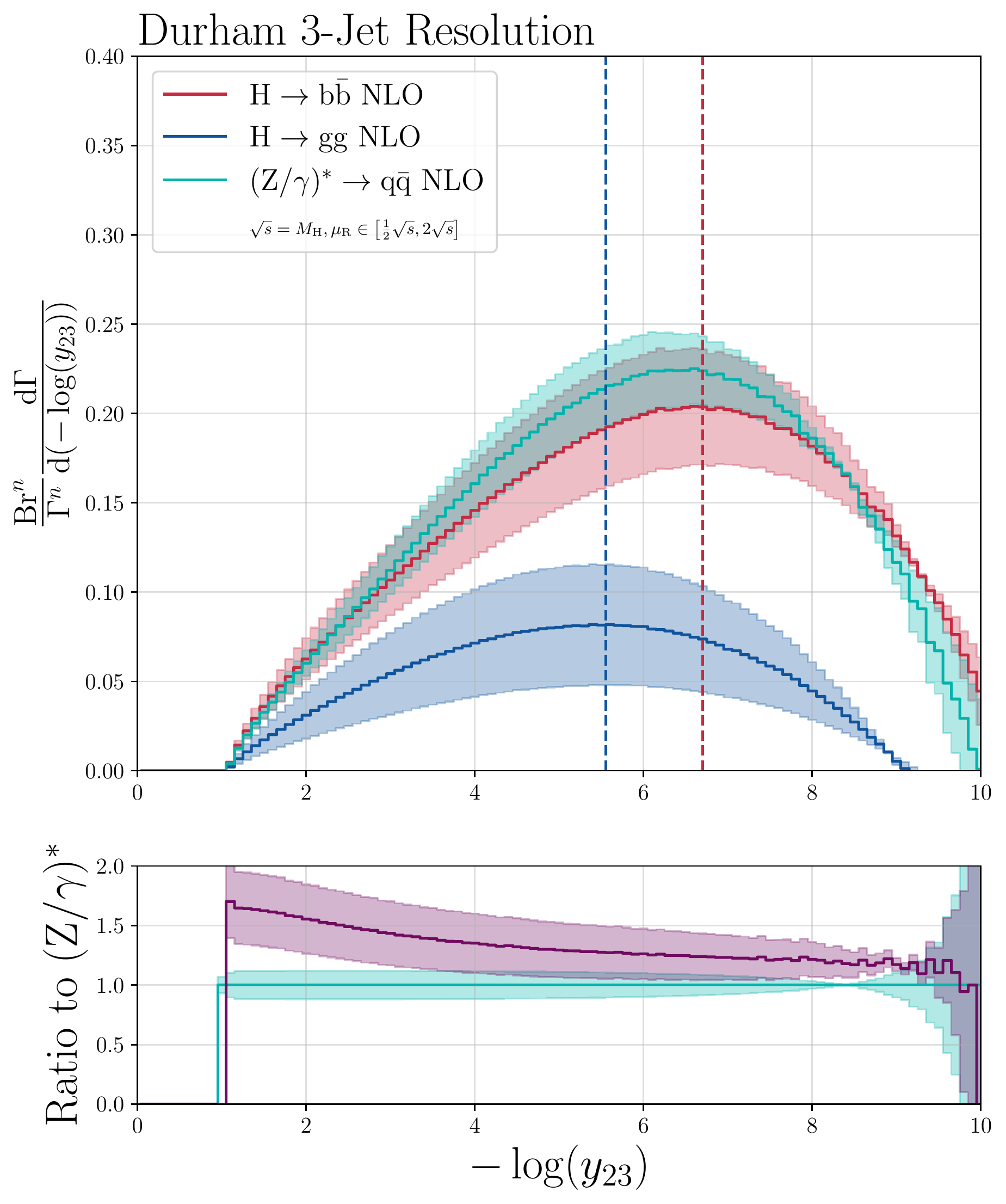}
  \caption{Comparison of event-shape distributions in hadronic Higgs decays and off-shell photon/$\PZ$ decays. In the upper panel, the maxima of the $\PH\to\Pqb\Paqb$ and $\PH\to\Pg\Pg$ distributions are marked with dashed red and blue lines, respectively (see text). In the lower panel, the ratio of the Higgs-decay result summed over the two categories is shown with respect to the respective $(\PZ/\Pgg)^*$-decay results.}
  \label{fig:cmp}
\end{figure}

To further gauge the discriminatory power of event shapes, in \cref{fig:cmp} we compare the event-shape distributions in hadronic Higgs decays to the distributions associated to off-shell photon/$\PZ$ decays as originally implemented in \eerad \cite{Gehrmann-DeRidder:2014hxk,Gehrmann-DeRidder:2007vsv,Gehrmann-DeRidder:2009fgd}.
To this end, we choose the same setup as in \cref{subsec:setup}, with a centre-of-mass energy of $\sqrt{s} = M_\PH$. The results are shown in \cref{fig:cmp}, consisting of two panels: the upper panel contains the $\PH\to\Pg\Pg$, $\PH\to\Pqb\Paqb$, and $(\PZ/\Pgg)^*\to\Pq\Paq$ distributions at \NLO, while the lower panel shows the ratio of the sum of the two Higgs-decay distributions to the off-shell photon/$\PZ$ distributions at \NLO level. We again show $\PH\to\Pqb\Paqb$ results in red and $\PH\to\Pg\Pg$ results in blue; in addition, results for $(\PZ/\Pgg)^*\to\Pq\Paq$ distributions are shown in green and inclusive $\PH$-decay distributions (\ie the sum of the two categories) are plotted in purple. 

Notwithstanding normalisation differences, the $(\PZ/\Pgg)^*\to \Pq\Paq$ distributions resemble closely the $\PH\to\Pqb\Paqb$ ones, with some small shape differences. Compared to the shape difference to the distributions in the $\PH\to\Pg\Pg$ category, however, these are only minor. The exception here seems to be the $C$-parameter, where the difference between the quark-induced Higgs decay and the off-shell photon/$\PZ$ decay is more pronounced.
Nevertheless, all quark-induced $(\PZ/\Pgg)^* \to\Pq\Paq$ and $\PH\to\Pqb\Paqb$ distributions peak at roughly equal event-shape values, whereas the peak is considerably shifted for the gluon-induced channel $\PH\to\Pg\Pg$. This is visible from the upper panel of \cref{fig:cmp}, where the maxima in the Higgs-decay distributions have been marked by coloured dashed lines.
This observation is in line with expectations, as higher-order corrections are expected in the region towards the right-hand side of the plots and resummation of large logarithms is necessary in the two-jet limit on the left edge of the linearly binned histograms.

In particular, in the logarithmically binned $y_{23}^\mathrm{D}$ distribution, rather good agreement in the shape of the two quark-induced distributions is found in the region $\log y_{23}^\mathrm{D} \geq -4$, with some deviations observed in the region $\log y_{23}^\mathrm{D} < -4$, with a slight shift of the maximum of the $(\PZ/\Pgg)^*\to\Pq\Paq$ distribution towards larger clustering scales.
The $\PH\to\Pg\Pg$ distribution, on the other hand, is visibly more narrow with a peak located much closer to the hard phase-space region.

The ratio of the inclusive hadronic Higgs-decay distributions, given by the sum of the individual $\PH\to\Pg\Pg$ and $\PH\to\Pqb\Paqb$ distributions, to the off-shell $(\PZ/\Pgg)$-decay result is shown as a dark purple line in the ratio panels of the figures in \cref{fig:cmp}. In all six event-shape distributions, this ratio shows a clear trend, namely that it is an almost linear distribution up to the three-parton kinematical limit. In particular, the combined Higgs-decay distributions all show a significantly higher rate towards the multi-jet limit. In the case of the Durham three-jet resolution, shown in the lower right-hand pane in \cref{fig:cmp}, this straight-forwardly translates into the presence of larger clustering scales for hadronic Higgs decays than for hadronic $\PZ/\Pgg$-decays.

It is also worth noting that scale uncertainties in the Higgs-decay results are generally larger than for the $(\PZ/\Pgg)^*$ decay. 
This is partly due to the additional branching-ratio prefactors $\mathrm{Br}^n$ in \cref{eq:dist}, which need to be taken into account in the scale-variation prescription to ensure consistency with the $\Gamma^n$ normalisation. While a variation of the $\PH\to\Pg\Pg$ and $\PH\to\Pqb\Paqb$ branching ratios introduces notable shifts in the relative size of the distributions, this additional shift is absent for off-shell photon/$\PZ$ decays, where only a single decay category contributes.
Note further that the scale uncertainties on the event shapes related to the $\PH\to\Pg\Pg$ category are already larger per se, due to their larger \NLO corrections, as already mentioned in \ref{subsec:Higgs_only}.

Most importantly, the comparison to results in $(\PZ/\Pgg)^*\to\Pq\Paq$ decays confirms our assessment that visible shape differences in the intermediate regions of the $\PH\to\Pqb\Paqb$ and $\PH\to\Pg\Pg$ distributions are primarily rooted in the distinct quark and gluon radiation patterns and may as such assist in the discrimination of Higgs decays to heavy quarks and gluons.

\section{Summary and Outlook}
\label{sec:conclusion}
In this paper, we have, for the first time, presented a full-fledged \NLO calculation of all six classical event-shape observables in hadronic Higgs decays involving three parton final states at Born level. The latter supplements an earlier \NLO calculation of the thrust observable (as presented in \cite{Gao:2019mlt}), by the $C$-parameter, the total and wide jet broadening, the heavy-jet mass, and the Durham three-jet resolution event shapes.

Using the antenna-subtraction framework, we have extended the existing \eerad \NNLO parton-level Monte Carlo generator to compute hadronic Higgs-decay event shapes at \NLO level. We have shown that all Higgs event-shape distributions have large \NLO corrections, with more important corrections found in the di-gluon induced case. Specifically, differential $K$-factors lie between $1.07-1.55$ in the $\PH\to\Pqb\Paqb$ decay mode, while being around $2.58-4.13$ in the $\PH\to\Pg\Pg$ mode, with different shapes and magnitudes for different event-shape distributions.
We demonstrated that at \NLO, different event-shape observables develop different shapes between the two Higgs decay categories and discussed the resulting applicability to discrimination purposes. In particular, for all observables, at \LO, we find that the ratio is rather flat around $17.5\%-18\%$, while at \NLO, the ratio is amplified both is size and shape and lies around $30\%-50\%$.
For the specific cases, we conclude that the $C$-parameter, the wide jet broadening, and the heavy-jet mass have no or very limited potential as Higgs decay class discriminators, while the thrust and the total jet broadening can be considered as suitable candidates for this purpose. Of all six event shapes studied in this work, the Durham three-jet resolution shows the best potential to accomplish this disentanglement, due to a notable difference in the clustering scales between quark- and gluon-induced jets.
We have verified our assessment by comparing event shapes in hadronic Higgs decays to event shapes in hadronic off-shell photon/$\PZ$ decays. In all event-shape distributions, this comparison indicated a clear shift of the peaks away from the two-jet limit for gluon-induced decays.

We observe large scale uncertainties at \NLO, suggesting potentially sizeable \NNLO corrections specially in the $\PH\to\Pg\Pg$ category.
The inclusion of second-order QCD corrections in the present computation leading to an extension of the validity of the theoretical predictions for Higgs decays to \NNLO accuracy as well as the combination with Higgs production at lepton and hadron colliders is facilitated by the flexibility of the current implementation. 

Since fixed-order calculations of event-shape observables are trustworthy only in a limited phase-space region, it is furthermore expedient to consider the resummation of logarithmically enhanced contributions, be it via the combination with analytic calculations or the matching to parton showers, \cf \cite{Frixione:2002ik,Nason:2004rx}. While the former gives an accurate account of the enhanced (next-to-)next-to-leading logarithmic contributions, the latter allows for particle-level corrections such as hadronisation. Both of these possible extensions are left for future work. 

\acknowledgments
We thank John Campbell and Imre Majer for helpful comments on the implementations of Higgs-decay one-loop matrix elements in \MCFM and \NNLOJET, respectively, and Thomas Gehrmann for useful discussions.
This research is supported by the Swiss National Science Foundation (SNF) under contract 200021-197130 
and by the Swiss National Supercomputing Centre (CSCS) under project ID ETH5f.

\bibliography{bibliography.bib,aux-bib.bib}

\end{document}